\documentclass[journal]{IEEEtran}
\usepackage{stfloats}
\usepackage{makecell}
\usepackage{graphicx}
\usepackage[cmex10]{amsmath}
\usepackage{cases}
\usepackage[tight,footnotesize]{subfigure}
\usepackage{amsthm} 
\usepackage{cite}
\usepackage{citesort}
\usepackage{amssymb}
\allowdisplaybreaks[4]
\usepackage{algorithm}
\usepackage{algorithmic}
\usepackage{multirow}
\usepackage{amsmath}
\usepackage{xcolor}
\usepackage{CJK}
\usepackage{subeqnarray}
\usepackage{cases}
\usepackage{enumerate}

\usepackage{stfloats}
\usepackage{multirow}

\newtheorem{lemma}{\hskip\parindent\bf{Lemma}}

\ifCLASSINFOpdf
\else
\fi

\begin{document}

\title{Reconfigurable Intelligent Surface Aided Mobile Edge Computing: From Optimization-Based to Location-Only Learning-Based Solutions}

\author{Xiaoyan Hu,~\IEEEmembership{Member,~IEEE,} Christos Masouros,~\IEEEmembership{Senior Member,~IEEE,} Kai-Kit~Wong,~\IEEEmembership{Fellow,~IEEE}

\thanks{
This work will be presented in part at the  International Conference on Communications (ICC), Virtual/Montreal, Canada, June 2021 \cite{ICC2021_X.Hu_Removing}.} 
\thanks{X. Hu, C. Masouros and K.-K. Wong are with the Department of Electronic and Electrical Engineering, University College London, London  WC1E 7JE, UK (Email: $\{\rm xiaoyan.hu, c.masouros,kai\text{-}kit.wong\}@ucl.ac.uk$).}
}
\maketitle
\vspace{-1.5cm}
\begin{abstract}
In this paper, we explore optimization-based and data-driven solutions in a reconfigurable intelligent surface (RIS)-aided multi-user mobile edge computing (MEC) system, where the user equipment (UEs) can partially offload their computation tasks to the access point (AP). We aim at maximizing the total completed task-input bits (TCTB) of all UEs with limited energy budgets during a given time slot, through jointly optimizing the RIS reflecting coefficients, the AP's receive beamforming vectors, and the UEs' energy partition strategies for local computing and offloading. A three-step block coordinate descending (BCD) algorithm is first proposed to effectively solve the non-convex TCTB maximization problem with guaranteed convergence. In order to reduce the computational complexity and facilitate lightweight online implementation of the optimization algorithm, we further construct two deep learning architectures. The first one takes channel state information (CSI) as input, while the second one exploits the UEs' locations only  for online inference. The two data-driven approaches are trained using data samples generated by the BCD algorithm via supervised learning. Our simulation results reveal a close match between the performance of the optimization-based BCD algorithm and the low-complexity learning-based architectures, all with superior performance to existing schemes in both cases with perfect and imperfect input features. Importantly, the location-only deep learning method is shown to offer a particularly practical and robust solution alleviating the need for CSI estimation and feedback when line-of-sight (LoS) direct links exist between UEs and the AP. 
\end{abstract}
\begin{IEEEkeywords}
Mobile edge computing, reconfigurable intelligent surface, receive beamforming, energy partition, deep learning. 
\end{IEEEkeywords}

\IEEEpeerreviewmaketitle

\vspace{-6mm}
\section{Introduction}\label{sec:Introduction}
\subsection{Motivations and Prior Works}\label{sec:PW}

The increasing data rates provided by 5G and beyond technologies, together with the proliferation of Internet-of-things (IoT) devices, have recently given rise to massive connectivity communications.
Accompanied by a wide range of emerging computation-intensive applications, the computing and processing demands for user equipment (UEs), e.g., smart devices and IoT sensors,  are  growing unprecedentedly. In order to liberate the resource-limited UEs from heavy computation workloads and provide them with high-performance low-latency computing services, mobile edge computing (MEC) promotes to use cloud computing capabilities at the edge of mobile networks through integrating MEC servers at the wireless access points (APs) \cite{S_Y.Mao17ASurve}. Hence, UEs' computation-intensive tasks can be offloaded and completed at the adjacent APs with less cost, energy and time.

Extensive works have contributed to the performance enhancement of applying MEC in various wireless networks, either improving the energy efficiency or reducing the  execution latency through jointly optimizing the radio and computational resources \cite{S.Sardellitti.15Joint,J_X.Chen2016Efficient,J_C.You17Energy,J_X.Hu18Wireless,J_T.X.Tran19Joint,TWC_X.Hu2020Edge}. 
A multicell MEC system was considered in \cite{S.Sardellitti.15Joint}, where users' energy consumption was minimized through joint resource allocation.
Later in \cite{J_X.Chen2016Efficient}, a game-theoretic algorithm was proposed to maximize the cell load as well as minimize the cost of time and energy.
The offloading priority function was defined in \cite{J_C.You17Energy} to show the relationship between the offloading strategy and resource allocation.
A wireless powered MEC system was investigated in \cite{J_X.Hu18Wireless}, where user cooperation was utilized to counteract the double-near-far effect.
Work in \cite{J_T.X.Tran19Joint} addressed the joint resource allocation of a multi-user multi-server MEC scenario to maximize the system utility.
The complementary benefits of edge and central cloud computing were studied in a two-tier heterogeneous cloud computing network in \cite{TWC_X.Hu2020Edge}. 

In order to further enhance the uplink offloading performance of the resource-limited UEs, great attentions have been drawn to the technology of reconfigurable intelligent surface (RIS) recently, due to its advantages of low cost, easy deployment, fine-grained passive beamforming and directional signal enhancement or interference nulling \cite{Acc_E.Basar2019Wireless,CM_Q.Wu2020,TWC_C.Huang2019Reconfigurable}. Through controlling the reflecting elements on the surface, RISs can be reconfigured to provide a more favourable wireless propagation environment for communications. Clearly, leveraging RISs into MEC systems is a cost-effective and environment-friendly way to facilitate UEs' computation offloading.

Several pioneer RIS-aided MEC works have been done  to explore the potential benefits of utilizing RISs in MEC systems \cite{JSAC_T.Bai2020Latency,ArXiv_S.Hua2019Reconfigurable,ArXiv_Y.Liu2020Intelligent,ArXiv_T.Bai2020Resource}.
A multi-user RIS-aided MEC system was considered in \cite{JSAC_T.Bai2020Latency}, where the execution latency was minimized with joint optimization on resource allocation and RIS coefficients design in an iterative way. It was verified that significant performance improvement can be attained compared to the case without RIS.
The advantages of RIS in directional beamforming were exploited for both uplink task offloading and downlink results downloading in \cite{ArXiv_S.Hua2019Reconfigurable}, where the power minimization problem was solved with an iterative block-structured algorithm.
Similarly, both uplink and downlink transmissions were considered in the RIS-aided MEC work \cite{ArXiv_Y.Liu2020Intelligent}, while the system utility was maximized iteratively to reduce the cost of energy and time.
Later in \cite{ArXiv_T.Bai2020Resource}, RIS was used in a wireless powered MEC system, and the energy consumption was minimized through a two-step iterative method.

For RIS-aided MEC systems, the formulated performance enhancement problems are typically non-convex with coupled optimization variables. Hence, iterative algorithms are usually necessary for jointly optimizing the radio and computational resource allocation as well as the RIS coefficients design. It is true that iterative algorithms may be capable of providing near-optimal solution even with guaranteed convergence, but they have very high computational complexity which require long execution time and thus may hinder their utilization in practical networks.
To tackle this issue, deep learning architectures provide a promising way to achieve lightweight online implementation via offline training \cite{J_K.Hornik1989Multilayer,B_I.Goodfellow2016Deep,TCCN_T.OShea2017AnIntroduction}.

Note that deep learning methods have been investigated in some MEC systems to simplify the optimization algorithm or fulfill online implementations \cite{Proceed_J.Chen2019Deep,Springer_L.Huang2018Distributed,TETP_J.Wang2019Smart,TMC_L.Huang2019Deep}.
A deep learning-based offloading strategy was designed in \cite{Springer_L.Huang2018Distributed} to minimize the  weighted energy consumption and latency. The deep reinforcement learning (DRL) was leveraged in  \cite{TETP_J.Wang2019Smart} for smart resource allocation of a software defined network (SDN)-enabled MEC architecture, also in \cite{TMC_L.Huang2019Deep} for online offloading decisions and resource allocation of a wireless powered MEC system.
Recently, the DRL was also used in the RIS-aided architectures to enhance the security \cite{TWC_H.Yang2020Deep} and maximize the sum rate of the downlink communications \cite{JSAC_C.Huang2020Reconfigurable}.
In \cite{WCL_A.M.Elbir2020DeepChannel}, a convolutional neural network (CNN) was constructed  for channel estimation of a large RIS-aided millimeter-wave (mm-wave) communication system.


\vspace{-2mm}
\subsection{Our Contributions}\label{sec:Contributions}
As per the above literature, deep learning approaches are promising to offer low-complexity solutions for the traditional MEC-related systems or RIS-aided downlink communication architectures.
However, the potentials of deep learning methods in simplifying the optimization algorithms of complex RIS-aided MEC systems have not been explored in the existing literature.
In this paper, a multi-user RIS-aided MEC architecture with multiplexing computation offloading is considered, where the RIS is installed to constructively control the interference and enhance the overall performance of UEs. We not only propose an iterative optimization algorithm to efficiently solve the formulated problem with guaranteed convergence, but also construct two deep learning architectures to facilitate online implementations of the proposed algorithm with significantly reduced complexity. 
To the best of our knowledge, this is the first work that leverages the data-driven approach in the RIS-aided MEC systems. Also, the proposed optimization-based algorithm is used to train the deep neural networks (DNNs) for efficient optimization and lightweight online implementation.

Our main contributions are summarized as follows:
\begin{itemize}
  \item \textbf{\em  A RIS-aided MEC architecture with uplink multiplexing offloading is leveraged to enhance the performance for maximizing the total completed task-input bits (TCTB) of all the resource-limited UEs.}
   Partial computation offloading is adopted for the RIS-aided MEC system in a multiplexing way, where the UEs can partially offload their computation task-input bits simultaneously. %
   We aim at maximizing the TCTB of all the UEs with limited energy supply budgets during a given time slot,  which  maximizes the computation efficiency in both time and energy, through jointly optimizing the RIS reflecting coefficients, the AP's receive beamforming vectors, and the UEs' energy partition strategies for local computing and computation offloading.
   The utilization of RIS is capable to enhance the performance of maximizing  TCTB by constructively  reconfiguring favourable propagations for all the UEs.

  \item \textbf{\em The RIS reflecting coefficients design, receive beamforming design, and energy partition optimization for maximizing the TCTB are effectively addressed through a three-step  block coordinate descending (BCD) optimization algorithm with guaranteed convergence.}
      The non-convex property and strongly coupled optimization variables of the formulated TCTB maximization problem make it difficult to obtain the global optimal solution.
      To address this issue, we propose a three-step BCD optimization algorithm to effectively separate the coupling and solve the problem by addressing three sub-problems iteratively. The DC  (difference of convex functions) programming method is leveraged to solve the first and third sub-problems respectively for RIS reflecting coefficients design and receive beamforming design with guaranteed convergence, while the optimal solution of the second sub-problem for receive beamforming design is obtained  via eigenvalue decompositions.

  \item \textbf{\em A CSI-based deep learning architecture with DNN-CSI is constructed to facilitate the online implementation of the proposed optimization-based BCD algorithm with significantly reduced  complexity.} Supervised learning is adopted to train the DNN-CSI using the data samples generated by the proposed BCD algorithm. It is shown that this CSI-based data-driven method can sufficiently capture the mapping of the BCD algorithm to the optimization solution, with lightweight inference complexity. Satisfactory and stable performance can be achieved in both scenarios without and with strong line-of-sight (LoS) direct links between the UEs and the AP.

  \item \textbf{\em A location-only deep learning architecture is further constructed that can effectively predict the solution of the proposed BCD algorithm without the need for pilot channel estimation and feedback during online inference.}
    This data-driven method is also based on supervised learning, and it performs well when LoS direct links are available for UEs. 
    The complexity of both training and testing can be greatly reduced compared with the CSI-based learning architecture since only UEs' locations are needed as the input feature. Thus, more lightweight online implementation can be achieved.

\item \textbf{\em   High effectiveness and robustness of the CSI-based and the location-only deep learning architectures  to the uncertainty of the input CSI and UEs' locations are validated.}
    The scenarios with corrupted input features to the two proposed deep learning architectures are considered to validate their effectiveness and robustness.
\end{itemize}

Simulation results are presented to evaluate the performance of the proposed BCD optimization algorithm and the two deep learning architectures.
It is confirmed that the proposed BCD algorithm highly outperforms the other three traditional benchmarks.
In addition, the CSI-based deep learning architecture can always approach the performance of the BCD algorithm in both scenarios without and with LoS direct links between the UEs and the AP.
It is noticeable that the location-only deep learning architecture can replace the CSI-based architecture to provide a satisfactory data-driven solution in the scenario with LoS direct links, with much less required overheads.
Besides, it is shown that the constructed two deep learning architectures can effectively emulating the proposed BCD algorithm even in the more practical scenarios with  uncertainty in the input information of CSI and UEs' locations.

The rest of this paper is organized as follows. Section~\ref{sec:system} introduces the considered system model and presents the corresponding  problem formulation. The BCD algorithm is proposed in Section~\ref{algorithm_design} to provide an optimization solution to the formulated problem, while two deep learning architectures are shown in Section \ref{sec:SL} to offer the learning-based \mbox{solutions}. The implementation setting and complexity reduction of the deep learning approaches are discussed in Section \ref{sec:DL_Comparison}. Section~\ref{sec:simulation} provides the simulation results, and we conclude our paper in Section~\ref{sec:conclusion}.

{\em Notations}---In this paper, the upper and lower
case bold symbols denote matrices and vectors, respectively. The notations $(\cdot)^\mathrm{T}$ and $(\cdot)^\mathrm{H}$ represent transpose and conjugate transpose for vectors or matrices.   
In addition, $\otimes$ denotes the Kronecker product.
$\mathrm{Tr}\left\{\mathbf{A}\right\}$ is the trace of square matrix $\mathbf{A}$. Also,
${\rm eig}\left\{\mathbf{A}\right\}$ denotes the set of all the eigenvalues of $\mathbf{A}$, and ${\rm eigvec}\left\{\cdot\right\}$ gives the eigenvector for a given eigenvalue of $\mathbf{A}$.
$\nabla_{\mathbf{X}}f(\mathbf{X})$ denotes the Jacobian matrix of function $f(\mathbf{X})$ with respect to (w.r.t.) the matrix $\mathbf{X}$, and $\partial_{\mathbf{X}}g(\mathbf{X})$ denotes a subgradient of function $g(\mathbf{X})$ w.r.t. $\mathbf{X}$. $\langle\mathbf{X}_1,\mathbf{X}_2\rangle\triangleq\mathfrak{R}\{\mathrm{tr}(\mathbf{X}_1^\mathrm{H}\mathbf{X}_2)\}$, where $\mathfrak{R}\{\cdot\}$ is the real-value operator. Finally, $\mathrm{diag}\{\mathbf{x}\}$ is the diagonal matrix formed by the elements of vector $\mathbf{x}$.
\vspace{-2mm}
\section{System Model and Problem Formulation}\label{sec:system}
\begin{figure}[tbp]
  \centering
  \includegraphics[scale=0.45]{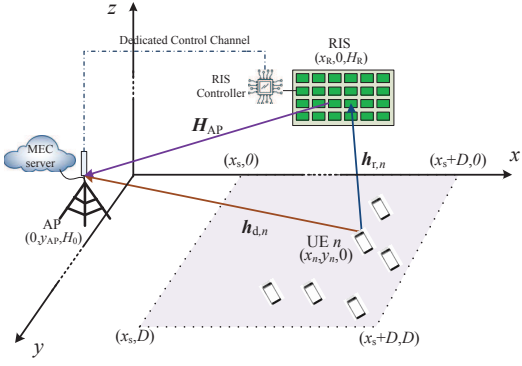}\\
  \vspace{-2mm}
  \caption{\footnotesize{An illustration of the RIS-aided MEC architecture, where a $K$-element RIS assists the multiplexing computation offloading of $N$ UEs. The phase shifts of the RIS elements can be adaptively adjusted by the AP through the dedicated control channel, so as to refine the signal propagations.}}\label{fig:system_model}
\end{figure}

We consider a RIS-aided MEC system as shown in Fig. \ref{fig:system_model}, which consists of $N$ single-antenna ground UEs, one RIS with $K$ reflecting elements, and one $M$-antenna AP. The RIS can be flexibly installed on the surrounding building walls, and it is under the control of the AP through a wireless controller to dynamically adjust the phase shift of each reflecting element.
By choosing a desirable location of the RIS, it is possible  to achieve LoS connections between the RIS and the AP as well as the UEs within a certain area.

\vspace{-4mm}
\subsection{Completed Task-Input Bits (CTB) with Partial Offloading}\label{sec:PartialOffloading} 
Each UE $n\in \mathcal{N}=\{1,2,\dots,N\}$ has intensive computation task-input bits (e.g., program codes and input parameters) to be dealt with, but with  a limited  energy budget dedicated for completing these task bits, denoted as $E_n$ in Joules (J).
A partial offloading mode is adopted to handle UEs' computation tasks, which are suitable to the augmented reality (AR) applications mentioned in \cite{S_Y.Mao17ASurve}. For these kinds of computation tasks, UEs' task-input bits can be arbitrarily divided to facilitate  parallel operations at UEs for local computing and offloading to the AP for remote computing.
Hence, the accounted computation energy consumption of each  UE includes that both used for local computing and computation offloading. The grid-powered AP is co-located with a powerful MEC server for helping UEs compute their offloaded tasks and it is also capable of downloading UEs' computation results, both in negligible time.\footnote{In this paper, we ignore the execution latency at the MEC server due to the fact that the MEC server is grid-power supplied and has a super-high computing capability, and thus the corresponding execution  latency  is negligible compared with that consumed at the UEs \cite{J_C.You17Energy,J_X.Hu18Wireless,TWC_X.Hu2020Edge,ArXiv_T.Bai2020Resource,TWC2018_S.Bi_Computation,TWC2020_F.Zhou_Computation,J_F.Zhou2018ComputationRM,TWC2020_X.Hu_Wireless,J_X.Hu19UAV}.
In addition, we assume that the UEs' computation results are with very small sizes, e.g., a few command bits, which can be ignored especially compared with their task-input bits, and thus the AP with a sufficient power supply can transmit the results back to UEs with negligible time \cite{J_C.You17Energy,J_X.Hu18Wireless,J_T.X.Tran19Joint,TWC_X.Hu2020Edge,JSAC_T.Bai2020Latency,ArXiv_Y.Liu2020Intelligent,ArXiv_T.Bai2020Resource,TWC2018_S.Bi_Computation,TWC2020_F.Zhou_Computation,J_F.Zhou2018ComputationRM}.}
We use $C_n$ to represent the amount of required computing resource, i.e., the number of CPU cycles, for completing 1-bit of UE $n$'s input data.
Our aim is to maximize the TCTB of all UEs each with a limited energy budget during a given time slot $T$, which is equivalent to maximizing the computation efficiency of the RIS-assisted MEC  system, including both energy efficiency and time efficiency referring to computation rate
\cite{TWC2018_S.Bi_Computation,TWC2020_F.Zhou_Computation,J_F.Zhou2018ComputationRM,TWC2020_X.Hu_Wireless}.

We first introduce a partition parameter $a_n \in [0,1]$ for UE $n\in\mathcal{N}$, and  $a_n E_n$ J of energy will be used for computation offloading while $(1-a_n) E_n$ J of energy will be used for local computing. In this case, the transmit power of UE $n$ for computation offloading is given as
\begin{align}\label{eq:p_n}
p_n=\frac{a_n E_n}{T}\triangleq a_n \widetilde{E}_n, \ \forall n\in\mathcal{N},
\end{align}
with $\widetilde{E}_n=E_n/T$ for $n\in\mathcal{N}$.

Let $s_n$ denote the task-input data-bearing signal transmitted by UE $n\in\mathcal{N}$ for computation offloading with $|s_n|=1$. Note that all the UEs with offloading requirements transmit their signals simultaneously in a multiplexing way within the given time slot, and thus we can express the corresponding received signal $\mathbf{y}\in\mathbb{C}^{M\times 1}$ at the AP as \cite{TWC_Q.Wu2019Intelligent} 
\begin{align}\label{eq:y_n}
\mathbf{y}=\sum\limits_{n{\rm{ = }}1}^{N}(\mathbf{H}_{\mathrm{AP}}\mathbf{\Phi} \mathbf{h}_{\mathrm{r},n}+\mathbf{h}_{\mathrm{d},n})\sqrt{p_n}s_n+\mathbf{n},
\end{align}
where  $\mathbf{h}_{\mathrm{d},n}\in\mathbb{C}^{M\times1}$ is the direct link between UE $n$ and the AP, $\mathbf{h}_{\mathrm{r},n}\in\mathbb{C}^{K\times1}$ indicates the relay channel between UE $n$ and the RIS, and $\mathbf{H}_{\mathrm{AP}}\in\mathbb{C}^{M\times K}$ represents the channel between RIS and the AP.
We assume that the channels  $\{\mathbf{h}_{\mathrm{d},n}\}$,  $\{\mathbf{h}_{\mathrm{r},n}\}$ and $\mathbf{H}_{\mathrm{AP}}$ are quasi-static within the given time slot.
Additionally, $\mathbf{\Phi}=\mathrm{diag}\{\boldsymbol{\phi}\}$ indicates the reflection-coefficient matrix of the RIS, where $\boldsymbol{\phi}=[\phi_1,\dots,\phi_K]^\mathrm{T}$ and $\phi_k=e^{j\theta_k}$ being the phase shift of the $k$-th reflecting element of the RIS with $\theta_k\in[0,2\pi]$ for $k\in\mathcal{K}=\{1,2,\dots,K\}$. Also, $\mathbf{n}\sim\mathcal{CN}(0,\sigma^2\mathbf{I}_M)$ is the the additive white Gaussian noise (AWGN) at the AP with $\sigma^2$ being the noise power. The linear beamforming strategy is then adopted at the AP to decode the UEs' transmit signals, and $\mathbf{w}_n\in\mathbb{C}^{M\times1}$ is the specific receive beamforming vector for UE $n$. Thus, the estimated signal for UE $n$ can be given as
\begin{align}\label{eq:s_cap_n}
\widehat{s}_n&=\mathbf{w}{_n^\mathrm{H}}\mathbf{y} \\ \nonumber
&=\mathbf{w}{_n^\mathrm{H}}\sum\limits_{n{\rm{ = }}1}^{N}(\mathbf{H}_{\mathrm{AP}}\mathbf{\Phi} \mathbf{h}_{\mathrm{r},n}+\mathbf{h}_{\mathrm{d},n})\sqrt{p_n}s_n+\mathbf{w}{_n^\mathrm{H}}\mathbf{n}, \ \forall n\in\mathcal{N}.
\end{align}
Based on the analysis above, we can obtain the uplink signal-to-interference-plus-noise ratio (SINR) for offloading UE $n$'s tasks as
\begin{align}\label{eq:SINR_n}
&\hspace{-1mm}\gamma_n(\mathbf{a},\mathbf{w}_n,\boldsymbol{\phi})=\\ \nonumber
&\hspace{-1mm}\frac{a_n\widetilde{E}_n|\mathbf{w}{_n^\mathrm{H}}(\mathbf{H}_{\mathrm{AP}}\mathbf{\Phi} \mathbf{h}_{\mathrm{r},n}+\mathbf{h}_{\mathrm{d},n})|^2}   {\sum\limits_{i=1, i\neq n}^{N} a_i\widetilde{E}_i|\mathbf{w}{_n^\mathrm{H}}(\mathbf{H}_{\mathrm{AP}}\mathbf{\Phi} \mathbf{h}_{\mathrm{r},i}+\mathbf{h}_{\mathrm{d},i})|^2 +\sigma^2\|\mathbf{w}{_n^\mathrm{H}}\|^2}, \forall n\in\mathcal{N},
\end{align}
where we denote an energy partition vector $\mathbf{a}=[a_1,\dots,a_N]$. Then, the CTB of UE  $n$ through computation offloading can be expressed as \cite{B_C.Shannon1948TheMathematical}
\begin{align}\label{eq:R_off_n}
\hspace{-2mm}
R{_n^{\mathrm{off}}}(\mathbf{a},\mathbf{w}_n,\boldsymbol{\phi})=BT\log_2(1+\gamma_n(\mathbf{a},\mathbf{w}_n,\boldsymbol{\phi})), \ \forall n\in\mathcal{N},
\end{align}
where $B$ is the given bandwidth shared with all the UEs.

As for the case of local computing, the dynamic voltage and frequency scaling (DVFS) technique is adopted at all the UEs for increasing the computation energy efficiency through adaptively controlling the CPU frequency used for computing \cite{J_Zhang13Energy}. Thus, the computation energy consumption of UE $n\in\mathcal{N}$ can be expressed  as $T\kappa_nf_n^3$, where $\kappa_n$ is the effective capacitance coefficient of UE $n$, and $f_n$ is the CPU frequency of its processing server. Also, we have $(1-a_n)E_n=T\kappa_nf_n^3$, and thus we can calculate $f_n$ as
\begin{align}\label{eq:f_n}
f_n=\sqrt[3]{\frac{(1-a_n)E_n}{T\kappa_n}}, \ \forall n\in\mathcal{N}.
\end{align}
Hence, the CTB of UE $n$ for local computing  can be given  as
\begin{align}\label{eq:R_loc_n}
R{_n^{\mathrm{loc}}}(a_n)=\frac{f_nT}{C_n}=\frac{T}{C_n}\sqrt[3]{\frac{ (1-a_n)\widetilde{E}_n }{ \kappa_n} }, \ \forall n\in\mathcal{N}.
\end{align}

\vspace{-4mm}
\subsection{Problem Formulation}\label{sec:problem}
We aim at maximizing the  TCTB of all the UEs with limited energy supply $\{E_n\}_{n\in\mathcal{N}}$ in the given time slot $T$, including their CTB through  both computation offloading and local computing, where the objective  TCTB is maximized through jointly optimizing the reflection coefficients in $\boldsymbol{\phi}$,  the receive beamforming vectors in $\mathbf{W}=[\mathbf{w}_1,\dots,\mathbf{w}_N]$, and the energy partition parameters in $\mathbf{a}$. As we mentioned before, maximizing the TCTB in this scenario can maximize the computation efficiency in both time and energy \cite{TWC2018_S.Bi_Computation,TWC2020_F.Zhou_Computation,J_F.Zhou2018ComputationRM,TWC2020_X.Hu_Wireless}.
The corresponding TCTB maximization  problem is formulated as problem (P0) given below
\begin{subequations}\label{eq:MSCT}
\begin{align}
({\rm P0}):
\underset{\mathbf{a},\mathbf{W},\boldsymbol{\phi}}{\max}
~~&\sum\limits_{n{\rm{ = }}1}^{N} \left(R{_n^{\mathrm{off}}}(\mathbf{a},\mathbf{w}_n,\boldsymbol{\phi})+
R{_n^{\mathrm{loc}}}(a_n)\right) \label{eq:MSCT_0}\\
\mbox{s.t.}~~~&a_n\in[0,1],  \ \forall n\in\mathcal{N},  \label{eq:MSCT_1}\\
&|\phi_n|=1, \ \forall n\in\mathcal{N},  \label{eq:MSCT_2}
\end{align}
\end{subequations}
which is a non-convex optimization problem since the optimization variables $\boldsymbol{\phi}$, $\mathbf{W}$, and $\mathbf{a}$ are strongly coupled. Hence, it is computationally difficult to find the global optimal solution of problem (P0), similar in \cite{JSAC_T.Bai2020Latency,ArXiv_S.Hua2019Reconfigurable,ArXiv_Y.Liu2020Intelligent,ArXiv_T.Bai2020Resource}. To address this issue, we propose a three-step BCD optimization algorithm to effectively separate the coupling among the optimization variables and  solve this problem iteratively with guaranteed convergence. 

\vspace{-2mm}
\section{BCD Optimization Algorithm Design}\label{algorithm_design}
The proposed BCD algorithm is operated in three major steps by solving three sub-problems iteratively. In the $\chi$-th ($\chi=1,2,\dots$) iteration, 
we first design the RIS reflecting coefficients in $\boldsymbol{\phi}$, with given $\mathbf{W}_{\chi-1}$ and $\mathbf{a}_{\chi-1}$ obtained in the previous iteration, and denote the solution as $\boldsymbol{\phi}_{\chi}$.
Then, with given $\mathbf{a}_{\chi-1}$ and the obtained $\boldsymbol{\phi}_{\chi}$, we show that the receive beamforming vectors in  $\mathbf{W}$ have closed-form optimal solutions, presented as  $\mathbf{W}_\chi$. We finally optimize the energy partition parameters in $\mathbf{a}$ with the obtained $\boldsymbol{\phi}_{\chi}$ and $\mathbf{W}_\chi$, indicating the solution as $\mathbf{a}_\chi$.
With the given initial $\mathbf{W}_{0}$ and $\mathbf{a}_{0}$, we can prove that the proposed three-step BCD algorithm can be performed  with guaranteed convergence. Next, we will demonstrate the details of the BCD optimization algorithm for solving the three sub-problems in the $\chi$-th iteration.

\vspace{-2mm}
\subsection{RIS Reflecting Coefficients Design}\label{RISReflectingCoefficientsDesign}
In the $\chi$-th iteration of the BCD algorithm, we first consider designing the RIS reflecting coefficients, i,e, $\boldsymbol{\phi}$, with given $\mathbf{W}=\mathbf{W}_{\chi-1}$ and $\mathbf{a}=\mathbf{a}_{\chi-1}$. With the given energy partition parameters in $\mathbf{a}$, we can equivalently obtain the UEs' transmit power as $p_n=a_n\widetilde{E}_n$. Then,  the RIS's reflecting coefficients design problem (P1) can be described as
\begin{subequations}\label{eq:MSCTPhi}
\begin{align}
({\rm P1}):
\underset{\boldsymbol{\phi}}{\max}
~~&\sum\limits_{n{\rm{ = }}1}^{N} \log_2(1+\gamma_n(\boldsymbol{\phi})) \label{eq:MSCTPhi_0}\\ 
\mbox{s.t.}~~~
&|\phi_k|=1, \ \forall k\in\mathcal{K},  \label{eq:MSCTPhi_1}
\end{align}
\end{subequations}
which is still non-convex and difficult to deal with directly. According to the expression of $\gamma_n(\boldsymbol{\phi})$ in \eqref{eq:SINR_n} for $n\in\mathcal{N}$, we can re-express $|\mathbf{w}{_n^\mathrm{H}}(\mathbf{H}_{\mathrm{AP}}\mathbf{\Phi} \mathbf{h}_{\mathrm{r},i}+\mathbf{h}_{\mathrm{d},i})|^2$  as
\begin{align} \label{eq:SINR_numerator}
&|\mathbf{w}{_n^\mathrm{H}}(\mathbf{H}_{\mathrm{AP}}\mathbf{\Phi} \mathbf{h}_{\mathrm{r},i}+\mathbf{h}_{\mathrm{d},i})|^2\\ \nonumber
=&|\mathbf{w}{_n^\mathrm{H}}\mathbf{H}_{\mathrm{AP}}
\mathrm{diag}(\mathbf{h}_{\mathrm{r},i})\boldsymbol{\phi}+\mathbf{w}{_n^\mathrm{H}}\mathbf{h}_{\mathrm{d},i} |^2 \\ \nonumber
=&|\mathbf{h}_{\mathrm{r},n,i}^{\mathrm{RIS}}\boldsymbol{\phi}+h_{\mathrm{d},n,i} |^2, \ \forall n,i\in\mathcal{N},
\end{align}
where $\mathbf{h}_{\mathrm{r},n,i}^{\mathrm{RIS}}=\mathbf{w}{_n^\mathrm{H}}\mathbf{H}_{\mathrm{AP}}
\mathrm{diag}(\mathbf{h}_{\mathrm{r},i}) \in\mathbb{C}^{1\times K}$ and $h_{\mathrm{d},n,i}=\mathbf{w}{_n^\mathrm{H}}\mathbf{h}_{\mathrm{d},i}$.
By defining a matrix $\mathbf{Q}_{n,i}\in\mathbb{C}^{(K+1)\times (K+1)}$  as
\begin{align}\label{Q_SDR}
\mathbf{Q}_{n,i}=\left[
\begin{aligned}
&(\mathbf{h}_{\mathrm{r},n,i}^{\mathrm{RIS}})^\mathrm{H}\mathbf{h}_{\mathrm{r},n,i}^{\mathrm{RIS}}, ~(\mathbf{h}_{\mathrm{r},n,i}^{\mathrm{RIS}})^\mathrm{H}h_{\mathrm{d},n,i} \\
&h{_{\mathrm{d},n,i}^\mathrm{H}}\mathbf{h}_{\mathrm{r},n,i}^{\mathrm{RIS}}, ~~~~~~~~~~~0
\end{aligned}\right], \ \forall n\in\mathcal{N},
\end{align}
and a vector $\boldsymbol{\widetilde{\phi}}=[\phi_1,\dots,\phi_K,\xi]^\mathrm{T} \in\mathbb{C}^{(K+1)\times 1}$ with an auxiliary scalar $\xi$, we can then re-express $|\mathbf{h}_{\mathrm{r},n,i}^{\mathrm{RIS}}\boldsymbol{\phi}+h_{\mathrm{d},n,i} |^2
=\boldsymbol{\widetilde{\phi}}^\mathrm{H}\mathbf{Q}_{n,i}\boldsymbol{\widetilde{\phi}}+|h_{\mathrm{d},n,i}|^2
=\mathrm{Tr}(\mathbf{Q}_{n,i}\mathbf{\Psi})+|h_{\mathrm{d},n,i}|^2$, 
where $\mathbf{\Psi}=\boldsymbol{\widetilde{\phi}}\boldsymbol{\widetilde{\phi}}^\mathrm{H}\in\mathbb{C}^{(K+1)\times (K+1)}$ is a positive semidefinite matrix (PSD) related to the RIS reflecting coefficients.

Note that each added item in the objective function, i.e., $\log_2(1+\gamma_n(\boldsymbol{\phi}) )$,  can be re-written as
\begin{align}\label{log_DC}
&\hspace{-1mm}\log_2(1+\gamma_n(\boldsymbol{\phi}) )=\log_2(1+\gamma_n(\boldsymbol{\widetilde{\phi}}) ) \\ \nonumber
&\hspace{-1mm}=\log_2\Bigg(\sum\limits_{j=1}^{N}p_j|\mathbf{w}{_n^\mathrm{H}}(\mathbf{H}_{\mathrm{AP}}\mathbf{\Phi} \mathbf{h}_{\mathrm{r},j}+\mathbf{h}_{\mathrm{d},j})|^2
+\sigma^2||\mathbf{w}{_n^\mathrm{H}}||^2\Bigg)\\ \nonumber
&\hspace{-1mm}-\log_2\Bigg(\sum\limits_{i=1,i\neq n}^{N}p_i|\mathbf{w}{_n^\mathrm{H}}(\mathbf{H}_{\mathrm{AP}}\mathbf{\Phi} \mathbf{h}_{\mathrm{r},i}+\mathbf{h}_{\mathrm{d},i})|^2 +\sigma^2||\mathbf{w}{_n^\mathrm{H}}||^2\Bigg)\\ \nonumber
&\hspace{-1mm}=\log_2\Bigg(\sum\limits_{j=1}^{N}p_j(\mathrm{Tr}(\mathbf{Q}_{n,j}\mathbf{\Psi})+|h_{\mathrm{d},n,j}|^2) +\sigma^2||\mathbf{w}{_n^\mathrm{H}}||^2\Bigg)\\ \nonumber
&\hspace{-1mm}-\log_2\Bigg(\sum\limits_{i=1,i\neq n}^{N} p_i(\mathrm{Tr}(\mathbf{Q}_{n,i}\mathbf{\Psi})
+|h_{\mathrm{d},n,i}|^2) +\sigma^2||\mathbf{w}{_n^\mathrm{H}}||^2\Bigg)\\ \nonumber
&\hspace{-1mm}\triangleq F_{1,n}(\mathbf{\Psi})-F_{2,n}(\mathbf{\Psi}), \ \forall n\in\mathcal{N},
\end{align}
where $F_{1,n}(\mathbf{\Psi})$ and $F_{2,n}(\mathbf{\Psi})$ are two concave functions w.r.t. $\mathbf{\Psi}$. Hence, the problem (P1) can be equivalently transformed into the following problem ($\widetilde{\mathrm{P}}$1)
\begin{subequations}\label{eq:MSCTPhi_DC}
\begin{align}
({\rm \widetilde{P}1}):
\underset{\mathbf{\Psi\succeq0}}{\max}
~~&\sum\limits_{n{\rm{ = }}1}^{N} F_{1,n}(\mathbf{\Psi})-F_{2,n}(\mathbf{\Psi}) \label{eq:MSCTPhi_DC_0}\\
\mbox{s.t.}~~~
&\mathbf{\Psi}_{k,k}=1, \ \forall k=1,2,\dots,K+1,  \label{eq:MSCTPhi_DC_1} \\
&\mathrm{rank}(\mathbf{\Psi})=1.
\end{align}
\end{subequations}
Even though the objective function in \eqref{eq:MSCTPhi_DC_0} and the rank-one constraint \eqref{eq:MSCTPhi_DC_1} make problem ($\widetilde{\mathrm{P}}$1) non-convex, it is easy to note that the objective function is a sum of differences of  concave functions.
Next, we will show that the DC programming \cite{B_P.Tao2005TheDC} can be leveraged to effectively address the issues of the objective function and the rank-one constraint.

As for the objective function, in the $(l+1)$-th ($l=0,1,\dots$) iteration of the DC programming, the second concave item, i.e., $F_{2,n}(\mathbf{\Psi})$ for $n\in\mathcal{N}$, can be approximated by its linear upper bound at the point $\mathbf{\Psi}^{(l)}$ (the solution obtained from the previous $l$-th iteration), which is given as
\begin{align}\label{F2_upper}
&F_{2,n}(\mathbf{\Psi})\leq \widehat{F}_{2,n}(\mathbf{\Psi};\mathbf{\Psi}^{(l)})=F_{2,n}(\mathbf{\Psi}^{(l)})+\\ \nonumber
&\frac{ \sum\limits_{i=1,i\neq n}^{N}p_i \left\langle(\mathbf{\Psi}-\mathbf{\Psi}^{(l)}),\nabla_{ \mathbf{\Psi} } \mathrm{Tr}(\mathbf{Q}_{n,i}\mathbf{\Psi})|_{\mathbf{\Psi}=\mathbf{\Psi}^{(l)}} \right\rangle }
{ \ln2\bigg( \sum\limits_{i=1,i\neq n}^{N} p_i(\mathrm{Tr}(\mathbf{Q}_{n,i}\mathbf{\Psi}^{(l)})
+|h_{\mathrm{d},n,i}|^2) +\sigma^2||\mathbf{w}{_n^\mathrm{H}}||^2 \bigg) },
\end{align}
where $\nabla_{ \mathbf{\Psi} } \mathrm{Tr}(\mathbf{Q}_{n,i}\mathbf{\Psi})|_{\mathbf{\Psi}=\mathbf{\Psi}^{(l)}}$ denotes the Jacobian matrix of $\mathrm{Tr}(\mathbf{Q}_{n,i}\mathbf{\Psi})$ w.r.t. $\mathbf{\Psi}$ at the point $\mathbf{\Psi}^{(l)}$, and it is easy to note that the equality holds  when $\mathbf{\Psi}=\mathbf{\Psi}^{(l)}$.

As for the rank-one constraint, it can be equivalentlly transformed into the following form
\begin{align}\label{RankOne}
\mathrm{Tr}(\mathbf{\Psi})-||\mathbf{\Psi}||_\mathrm{s}=0,
\end{align}
where $||\mathbf{\Psi}||_\mathrm{s}$ denotes the spectral norm of the PSD matrix $\mathbf{\Psi}$. It is noticeable that $\mathrm{Tr}(\mathbf{\Psi})=\sum{_{k=1}^{K+1}}\rho_k(\mathbf{\Psi})$ and $||\mathbf{\Psi}||_\mathrm{s}=\rho_1(\mathbf{\Psi})$, where $\rho_k(\mathbf{\Psi})$ indicates the $k$-th largest singular value of  $\mathbf{\Psi}$. Hence, the equality of $\mathrm{Tr}(\mathbf{\Psi})=||\mathbf{\Psi}||_\mathrm{s}$ holds when the rank-one constraint is satisfied with $\rho_1(\mathbf{\Psi})>0$ and $\rho_k(\mathbf{\Psi})=0$ for $k=2,\dots,K+1$, and vice versa.
Similarly, in the $(l+1)$-th iteration of the DC programming, a linear lower-bound of the convex item $||\mathbf{\Psi}||_\mathrm{s}$ at the point $\mathbf{\Psi}^{(l)}$ can be expressed as
\begin{align}\label{RankOne_DC}
||\mathbf{\Psi}||_\mathrm{s}&\geq ||\mathbf{\Psi}^{(l)}||_\mathrm{s}+ \left\langle(\mathbf{\Psi}-\mathbf{\Psi}^{(l)}), \partial_\mathbf{\Psi} ||\mathbf{\Psi}||_\mathrm{s}|_{\mathbf{\Psi}=\mathbf{\Psi}^{(l)}} \right\rangle
\\ \nonumber
&\triangleq \Upsilon(\mathbf{\Psi};\mathbf{\Psi}^{(l)}),
\end{align}
where $\partial_\mathbf{\Psi} ||\mathbf{\Psi}||_\mathrm{s}|_{\mathbf{\Psi}=\mathbf{\Psi}^{(l)}}$ is a subgradient of the spectral norm $||\mathbf{\Psi}||_\mathrm{s}$ w.r.t. $\mathbf{\Psi}$  at the point $\mathbf{\Psi}^{(l)}$, and the equality holds when $\mathbf{\Psi}=\mathbf{\Psi}^{(l)}$.
Note that one subgradient of $||\mathbf{\Psi}||_\mathrm{s}$ at point $\mathbf{\Psi}^{(l)}$ can be efficiently computed as $\mathbf{z}_1\mathbf{z}{_1^\mathrm{H}}$, where $\mathbf{z}_1$ is the vector corresponding to the largest singular value of $\mathbf{\Psi}^{(l)}$ \cite{TWC_K.Yang2020Federated}.

With the obtained linear lower bound of $||\mathbf{\Psi}||_\mathrm{s}$ in \eqref{RankOne_DC}, we can generate an approximate rank-one constraint of \eqref{RankOne}, which is shown as
\begin{align}\label{RankOne1}
\mathrm{Tr}(\mathbf{\Psi})-\Upsilon(\mathbf{\Psi};\mathbf{\Psi}^{(l)}) \leq \varepsilon_\Psi,
\end{align}
where $\varepsilon_\Psi$ is a positive threshold with very small value close to zero.
The approximated rank-one constraint can guarantee that
$0\leq\mathrm{Tr}(\mathbf{\Psi})-||\mathbf{\Psi}||_\mathrm{s}\leq\mathrm{Tr}(\mathbf{\Psi})-\Upsilon(\mathbf{\Psi};\mathbf{\Psi}^{(l)})\leq\varepsilon_\Psi$, and the rank-one constraint can be approached with an arbitrary accuracy by setting $\varepsilon_\Psi$  infinitely close to zero.

Hence, we can obtain an approximation problem of ($\widetilde{\mathrm{P}}$1) at the $(l+1)$-th iteration as
\begin{subequations}\label{eq:MSCTPhi_DCapp}
\begin{align}
({\rm P1.1}):
\underset{\mathbf{\Psi\succeq0}}{\max}
~~&\sum\limits_{n{\rm{ = }}1}^{N} F_{1,n}(\mathbf{\Psi})-\widehat{F}_{2,n}(\mathbf{\Psi};\mathbf{\Psi}^{(l)}) \label{eq:MSCTPhi_DCapp_0}\\
\mbox{s.t.}~~~
&\mathbf{\Psi}_{k,k}=1, \ \forall k=1,2,\dots,K+1,  \label{eq:MSCTPhi_DCapp_1} \\
&\mathrm{Tr}(\mathbf{\Psi})-\Upsilon(\mathbf{\Psi};\mathbf{\Psi}^{(l)})\leq \varepsilon_\Psi, \label{eq:MSCTPhi_DCapp_2}
\end{align}
\end{subequations}
which is a convex optimization problem and  can be readily solved by the existing convex solvers such as CVX \cite{M_Grant08CVX}, and the optimal solution can be obtained  as $\mathbf{\Psi}^{(l+1)}$. Through choosing $\mathbf{\Psi}^{(0)}=\mathbf{\Psi}_{\chi-1}=\widetilde{\boldsymbol{\phi}}_{\chi-1} \widetilde{\boldsymbol{\phi}}{_{\chi-1}^\mathrm{H}}$, it is easy to prove that the feasibility of problem (P1.1) in each iteration $l$ can always be guaranteed since $\mathbf{\Psi}^{(l-1)}$ is always a feasible solution. 

\begin{lemma}\label{lemma3}
The objective function of problem (P1) in \eqref{eq:MSCTPhi_DC_0} monotonically increases with the iteration index $l$ as
\begin{align}\label{eq:Increase_DC}
&F_{1,n}(\mathbf{\Psi}^{(l+1)})-F_{2,n}(\mathbf{\Psi}^{(l+1)}) \\ \nonumber
\overset{(a)}{\geq} &F_{1,n}(\mathbf{\Psi}^{(l+1)})-\widehat{F}_{2,n}(\mathbf{\Psi}^{(l+1)};\mathbf{\Psi}^{(l)}) \\ \nonumber
\overset{(b)}{\geq} &F_{1,n}(\mathbf{\Psi}^{(l)})-\widehat{F}_{2,n}(\mathbf{\Psi}^{(l)};\mathbf{\Psi}^{(l)})\\ \nonumber
=&F_{1,n}(\mathbf{\Psi}^{(l)})-F_{2,n}(\mathbf{\Psi}^{(l)}), \ \forall n\in\mathcal{N},
\end{align}
where $(a)$ comes from the inequality \eqref{F2_upper} and $(b)$ holds since $\mathbf{\Psi}^{(l)}$ is a feasible solution while $\mathbf{\Psi}^{(l+1)}$  is the optimal solution of problem (P1.1) in \eqref{eq:MSCTPhi_DCapp}. Also, the objective function of problem ($\widetilde{P}$1) is upper-bounded by the UEs' limited energy budgets. Hence, Problem ($\widetilde{P}$1)  in \eqref{eq:MSCTPhi_DC} as well as its equivalent form (P1) in \eqref{eq:MSCTPhi} can be solved through the DC programming method with guaranteed convergence \cite{B_P.Tao2005TheDC}. The final solution of $\mathbf{\Psi}$ at the convergence of the $(l+1)$-th iteration of the DC programming  is the solution  of the BCD algorithm at the $\chi$-th iteration, i.e., $\mathbf{\Psi}_{\chi}=\mathbf{\Psi}^{(l+1)}$.
\end{lemma}

With the obtained $\mathbf{\Psi}_{\chi}$, we can retrieve $\widetilde{\boldsymbol{\phi}}_\chi$ by decomposing $\mathbf{\Psi}_{\chi}
=\widetilde{\boldsymbol{\phi}}_\chi\widetilde{\boldsymbol{\phi}}{_\chi^\mathrm{H}}$ with denoting $\widetilde{\boldsymbol{\phi}}_\chi=[\boldsymbol{\phi}_{\chi,0},\xi_{\chi,0}]^\mathrm{T}$, and then it is easy to obtain the RIS reflecting coefficient vector at the $\chi$-th iteration of the BCD algorithm as $\boldsymbol{\phi}_\chi=\boldsymbol{\phi}_{\chi,0}/\xi_{\chi,0}$ and accordingly $\mathbf{\Phi}_\chi=\mathrm{diag}\{\boldsymbol{\phi}_\chi\}$.
In order to facilitate the following analysis of designing the algorithm, we define the effective UE-AP channels with given $\mathbf{\Phi}$ (or $\boldsymbol{\phi}$) as
\begin{align}\label{eq:h_n}
\mathbf{h}_n(\mathbf{\Phi})=\mathbf{H}_{\mathrm{AP}}\mathbf{\Phi} \mathbf{h}_{\mathrm{r},n}+\mathbf{h}_{\mathrm{d},n}, \ \forall n\in\mathcal{N}.
\end{align}

\subsection{Receive Beamforming Design}\label{ReceiveBeamformingDesign}
With given $\mathbf{a}=\mathbf{a}_{\chi-1}$ ($p_n=a_n\widetilde{E}_n$, $n\in\mathcal{N}$) and  $\mathbf{\Phi}=\mathbf{\Phi}_\chi$, the sub-problem for optimizing the AP's receive beamforming vextors for each UE, i,e, $\mathbf{w}_n$ for $n\in\mathcal{N}$, can be expressed as the following problem (P2)
\begin{align}\label{eq:MSCTW}
({\rm P2}):
\underset{\mathbf{W}}{\max}
~~&\sum\limits_{n{\rm{ = }}1}^{N} R{_n^{\mathrm{off}}}(\mathbf{w}_n),
\end{align}
which can be equivalently solved by addressing $N$ parallel sub-problems for each $n\in\mathcal{N}$ as
\begin{align} \label{eq:MSCTWn}
({\rm P2.1}):
\underset{\mathbf{w_n}}{\max}
~~& \gamma_n(\mathbf{w}_n)=\frac{\mathbf{w}{_n^\mathrm{H}}\mathbf{\Theta}_n \mathbf{w}_n}{\mathbf{w}{_n^\mathrm{H}}\mathbf{\Theta}_{-n} \mathbf{w}_n},
\end{align}
where $\mathbf{\Theta}_n =p_n {\mathbf{h}_{n}}({\mathbf{h}_{n}})^\mathrm{H}$ and $\mathbf{\Theta}_{-n}=\sum_{i=1, i\neq n}^{N} p_i{\mathbf{h}_{i}}({\mathbf{h}_{i}})^\mathrm{H}+\sigma^2\mathbf{I}_M$, with the effective channel $\{\mathbf{h}_n\}_{n\in\mathcal{N}}$ given in \eqref{eq:h_n}.

\begin{lemma}\label{lemma1}
It is easy to note that problem (P2.1) in \eqref{eq:MSCTWn} is a generalized eigenvector problem, and  its optimal solution $\mathbf{w}_n^*$ should be the  eigenvector corresponds to the largest eigenvalue of the matrix $\left(\mathbf{\Theta}_{-n}\right)^{-1} \mathbf{\Theta}_n$. Hence, the optimal $\mathbf{w}_n^*$ of problem (P2.1) for $n\in\mathcal{N}$ can be given as
\begin{align}\label{w_beam}
\mathbf{w}_n^*={\rm eigvec}\left\{\max\left\{{\rm eig}\{\left(\mathbf{\Theta}_{-n}\right)^{-1} \mathbf{\Theta}_n \}\right\}\right\}.
\end{align}
\end{lemma}
We then denote the receive beamforming matrix obtained at the $\chi$-th iteration of the BCD algorithm as $\mathbf{W}_\chi=[\mathbf{w}_1^*,\dots,\mathbf{w}_N^*]$, which is used in the following subsection.

\subsection{Energy Partition Optimization}\label{EnergyPartitionOptimization}
Here, the sub-problem (P3) for optimizing the energy partition parameters in $\mathbf{a}$ with given  $\mathbf{\Phi}=\mathbf{\Phi}_\chi$ and  $\mathbf{W}=\mathbf{W}_\chi$ is considered, which is given below
\begin{subequations}\label{eq:MSCTa}
\begin{align}
({\rm P3}):
\underset{\mathbf{a}}{\max}
~~&\sum\limits_{n{\rm{ = }}1}^{N} \left(R{_n^{\mathrm{off}}}(\mathbf{a})+
R{_n^{\mathrm{loc}}}(a_n)\right) \label{eq:MSCTa_0}\\
\mbox{s.t.}~~~&a_n\in[0,1],  \ \forall n\in\mathcal{N}.  \label{eq:MSCTa_1}
\end{align}
\end{subequations}
Note that problem (P3) is non-convex because of the non-concave items $\{R{_n^{\mathrm{off}}}(\mathbf{a})\}_{n\in\mathcal{N}}$ in the objective function \eqref{eq:MSCTa_0}. Actually, $R{_n^{\mathrm{off}}}(\mathbf{a})$ for $n\in\mathcal{N}$ can be re-expressed as the difference of two concave functions as follows
\begin{align}\label{eq:R_off_n_Difference}
&\hspace{-1mm}R{_n^{\mathrm{off}}}(\mathbf{a})\triangleq R{_{n,1}^{\mathrm{off}}}(\mathbf{a})-R{_{n,2}^{\mathrm{off}}}(\mathbf{a}_{-n})= \\ \nonumber
&\hspace{-1mm}BT\log_2\Bigg(\sum\limits_{j=1}^{N}a_j\widetilde{E}_j|\mathbf{w}{_n^\mathrm{H}}\mathbf{h}_{j}|^2
+\sigma^2||\mathbf{w}{_n^\mathrm{H}}||^2\Bigg)-\\ \nonumber
&\hspace{-1mm}BT\log_2\Bigg(\sum\limits_{i=1,i\neq n}^{N}a_i\widetilde{E}_i|\mathbf{w}{_n^\mathrm{H}}\mathbf{h}_{i}|^2 +\sigma^2||\mathbf{w}{_n^\mathrm{H}}||^2\Bigg),
\end{align}
where $\mathbf{a}_{-n}=[a_1,\dots,a_{n-1},a_{n+1},\dots,a_N]$.

Then the problem (P3)  can also be solved with the DC programming method, where the second concave function in \eqref{eq:R_off_n_Difference}, i.e.,   $R{_{n,2}^{\mathrm{off}}}(\mathbf{a}_{-n})$, can be substituted by its linear upper bound, so as to obtain a concave approximation of $R{_n^{\mathrm{off}}}(\mathbf{a})$. Assuming $\mathbf{a}^{(m)}$ is the solution obtained at the  $m$-th ($m=0,1,\dots$) iteration of the DC programming,  a linear upper bound of $R{_{n,2}^{\mathrm{off}}}(\mathbf{a}_{-n})$ at the point $\mathbf{a}^{(m)}$ can be obtained through the first-order Taylor series expansion  as
\begin{align}\label{eq:R_off_n2_upper}
&R{_{n,2}^{\mathrm{off}}}(\mathbf{a}{_{-n}})\leq \widehat{R}{_{n,2}^{\mathrm{off}}}(\mathbf{a}{_{-n}};\mathbf{a}{_{-n}^{(m)}})\\
&=R{_{n,2}^{\mathrm{off}}}(\mathbf{a}{_{-n}^{(m)}})+\sum\limits_{i=1,i\neq n}^{N}R{_{n,2,i}^{\mathrm{off}'}}(\mathbf{a}{_{-n}^{(m)}})*(a_i-a{_i^{(m)}}), \nonumber
\end{align}
where $R{_{n,2,i}^{\mathrm{off}'}}(\mathbf{a}{_{-n}^{(m)}})=\frac{BT}{\ln2}
\frac{ \widetilde{E}_i|\mathbf{w}{_n^\mathrm{H}}\mathbf{h}_{i}|^2 }
{ \sum\limits_{j=1,j\neq n}^{N}a_j^{(m)}\widetilde{E}_j |\mathbf{w}{_n^\mathrm{H}}\mathbf{h}_{j}|^2 +\sigma^2||\mathbf{w}{_n^\mathrm{H}}||^2 }$ is the first-order derivative of $R{_{n,2}^{\mathrm{off}}}(\mathbf{a}_{-n})$ w.r.t. $a_i$ at the point $\mathbf{a}{_{-n}^{(m)}}$. It is easy to note that the equality holds  when $\mathbf{a}{_{-n}}=\mathbf{a}{_{-n}^{(m)}}$.
At the $(m+1)$-th iteration of DC programming, we aim at maximizing the following approximation problem
\begin{subequations}\label{eq:MSCTaDC}
\begin{align}
\hspace{-4mm}({\rm P3.1}):
\underset{\mathbf{a}}{\max}
~&\sum\limits_{n{\rm{ = }}1}^{N} \left( R{_{n,1}^{\mathrm{off}}}(\mathbf{a})-\widehat{R}{_{n,2}^{\mathrm{off}}}(\mathbf{a}{_{-n}};\mathbf{a}{_{-n}^{(m)}})+
R{_n^{\mathrm{loc}}}(a_n)\right) \label{eq:MSCTaDC_0}\\
\hspace{-4mm}\mbox{s.t.}~~&a_n\in[0,1],  \ \forall n\in\mathcal{N},  \label{eq:MSCTaDC_1}
\end{align}
\end{subequations}
which is a convex  problem and can be easily solved by CVX \cite{M_Grant08CVX}. Through solving problem (P3.1) with CVX, the optimal solution, i.e., $\mathbf{a}^{(m+1)}$, can be finally obtained. 

\begin{lemma}\label{lemma2}
The objective function  of problem (P3) in \eqref{eq:MSCTa_0} is monotonic increasing w.r.t the iteration index $m$ as
\begin{align}\label{eq:IncreaseP2_DC}
 &R{_n^{\mathrm{off}}}( \mathbf{a}^{(m+1)} )+R{_n^{\mathrm{loc}}}(a_n^{(m+1)}) \\  \nonumber
\overset{(a)}{\geq} &R{_{n,1}^{\mathrm{off}}}( \mathbf{a}^{(m+1)} )-\widehat{R}{_{n,2}^{\mathrm{off}}}
( \mathbf{a}{_{-n}^{(m+1)}};\mathbf{a}{_{-n}^{(m)}})+R{_n^{\mathrm{loc}}}(a_n^{(m+1)}) \\  \nonumber
\overset{(b)}{\geq} &R{_{n,1}^{\mathrm{off}}}( \mathbf{a}^{(m)} )-\widehat{R}{_{n,2}^{\mathrm{off}}}
( \mathbf{a}{_{-n}^{(m)}};\mathbf{a}{_{-n}^{(m)}})+R{_n^{\mathrm{loc}}}(a_n^{(m)}) \\  \nonumber
= &R{_n^{\mathrm{off}}}( \mathbf{a}^{(m)} )+R{_n^{\mathrm{loc}}}(a_n^{(m)}),
\end{align}
where $(a)$ comes from the inequality \eqref{eq:R_off_n2_upper} and $(b)$ holds since $\mathbf{a}^{(m)}$ is a feasible solution while $\mathbf{a}^{(m+1)}$  is the optimal solution of problem (P3.1) in \eqref{eq:MSCTaDC}.
Besides, the objective \eqref{eq:MSCTa_0} is upper-bounded due to the limited energy supply of UEs. In summary, the convergence of the proposed DC programming method for solving problem (P3) in \eqref{eq:MSCTa} can be guaranteed \cite{B_P.Tao2005TheDC}. We can obtain the final solution of $\mathbf{a}$ at the $\chi$-th iteration of the BCD algorithm when the DC programming converges at the $(m+1)$-th iteration, which is denoted as $\mathbf{a}_{\chi}=[a_1^{(m+1)},\dots,a_N^{(m+1)}]$.
\end{lemma}

\subsection{Benchmark with Zero Forcing (ZF) Receive Beamforming}\label{ZF_Beamforming}
An important benchmark scheme of our proposed algorithm  is leveraging ZF receive beamforming at the AP, and thus the receive beamforming matrix obtained in Section \ref{ReceiveBeamformingDesign} should be replaced as $\mathbf{W}_\chi=[\mathbf{w}_{1,\mathrm{ZF}},\dots,\mathbf{w}_{N,\mathrm{ZF}}]=\mathbf{H}(\mathbf{H}^\mathrm{H}\mathbf{H})^{-1}$ in the $\chi$-th iteration of the proposed BCD algorithm, where $\mathbf{H}=[\mathbf{h}_1,\dots,\mathbf{h}_N] \in \mathbb{C}^{M\times N}$ is the compact matrix of the UEs' equivalent channels. For the case of $M\geq N$ with independent and identically distributed (i.i.d.) channels, the interference among the UEs can be eliminated, and thus the offloading rate of UE $n$ can be rewritten as
\begin{align}\label{eq:R_off_n_ZF}
\hspace{-2mm}
R{_{n,\mathrm{ZF}}^{\mathrm{off}}}(a_n)=BT\log_2\Bigg( 1+\frac{a_n\widetilde{E}_n}{\sigma^2||\mathbf{w}{_{n,\mathrm{ZF}}^\mathrm{H}}||^2} \Bigg), \ \forall n\in\mathcal{N},
\end{align}
which is concave w.r.t. $a_n$.
Then the problem (P3) in \eqref{eq:MSCTa} is reduced to a convex optimization problem, which can be optimally solved by CVX in a parallel fashion by addressing $N$ sub-problems for $n\in\mathcal{N}$ given below
\begin{subequations}\label{eq:MSCTa_UEn} 
\begin{align}
({\rm P3.2}):
\underset{a_n}{\max}
~~&\left(R{_{n,\mathrm{ZF}}^{\mathrm{off}}}(a_n)+
R{_n^{\mathrm{loc}}}(a_n)\right) \label{eq:MSCTa_UEn_0}\\
\mbox{s.t.}~~~&a_n\in[0,1].    \label{eq:MSCTa_UEn_1}
\end{align}
\end{subequations}
It is known that the ZF receive beamforming cannot effectively deal with the cases when $M<N$, while our proposed optimal solution in  Section \ref{ReceiveBeamformingDesign} can perform well even in these cases, which will be validated in the simulation results. 

\subsection{Algorithm, Convergence, and Complexity}
\begin{algorithm}[!htp]
\caption{Three-Step BCD Optimization Algorithm for Solving the Original TCTB Maximization Problem (P0)  }\label{algorithmic1} 
\begin{algorithmic}[1]
{\small{
\STATE \textbf{Input}  $T$, $N$, $M$, $K$,  $B$, $\{E_n, C_n, \kappa_n, \mathbf{h}_{\mathrm{d},n}, \mathbf{h}_{\mathrm{r},n}\}_{n\in\mathcal{N}}$,  $\mathbf{H}_{\mathrm{AP}}$, and the tolerant thresholds $\varepsilon$, $\varepsilon_1$, $\varepsilon_3$;
\STATE \textbf{Initialize} The iteration index $\chi=0$ and  $\boldsymbol{\phi}_{0}$, $\mathbf{W}_{0}$, $\mathbf{a}_{0}$;
\STATE \textbf{Repeat}
\STATE $\chi=\chi+1$; \\
\STATE~\textbf{Step 1:} \textbf{Initialize}  $l=0$, $\boldsymbol{\phi}^{(0)}=\boldsymbol{\phi}_{\chi-1}$, $\mathbf{W}=\mathbf{W}_{\chi-1}$, $\mathbf{a}=\mathbf{a}_{\chi-1}$;\\
\STATE~\textbf{Repeat 1}
\STATE~\quad a) Solve problem (P1.1) with CVX to obtain $\mathbf{\Psi}^{(l+1)}$;\\
\STATE~\quad b) Calculate  $R_1^{(l+1)}=\sum_{n{\rm{ = }}1}^{N} F_{1,n}(\mathbf{\Psi}^{(l+1)})-F_{2,n}(\mathbf{\Psi}^{(l+1)})$; \\
\STATE~\quad c) $l=l+1$;\\
\STATE~~\textbf{End Repeat 1} until convergence, i.e., $|R_1^{(l)}-R_1^{(l-1)}|<\varepsilon_1$ \\
      ~~($l>1$), and obtain $\mathbf{\Psi}_{\chi}=\mathbf{\Psi}^{(l)}$; Then we decompose $\mathbf{\Psi}_{\chi}
=\widetilde{\boldsymbol{\phi}}_\chi\widetilde{\boldsymbol{\phi}}{_\chi^\mathrm{H}}$ with denoting $\widetilde{\boldsymbol{\phi}}_\chi=[\boldsymbol{\phi}_{\chi,0},\xi_{\chi,0}]^\mathrm{T}$; Thus, we can obtain $\boldsymbol{\phi}_\chi=\boldsymbol{\phi}_{\chi,0}/\xi_{\chi,0}$ and $\mathbf{\Phi}_\chi=\mathrm{diag}\{\boldsymbol{\phi}_\chi\}$; \\
\STATE~\textbf{Step 2:} \textbf{Initialize} $\boldsymbol{\phi}=\boldsymbol{\phi}_{\chi}$ and $\mathbf{a}=\mathbf{a}_{\chi-1}$;
\STATE~Obtain $\mathbf{W}_\chi$ according to \textbf{Lemma} \ref{lemma1};
\STATE~\textbf{Step 3:}  \textbf{Initialize}  $m=0$, $\mathbf{a}^{(0)}=\mathbf{a}_{\chi-1}$, $\boldsymbol{\phi}=\boldsymbol{\phi}_{\chi}$, $\mathbf{W}=\mathbf{W}_\chi$;
\STATE~\textbf{Repeat 3}
\STATE~\quad a) Solve problem (P3.1) with CVX to obtain $\mathbf{a}^{(m+1)}$; \\
      ~\quad b) Calculate the TCTB at the $(m+1)$-th iteration, represented as $R_3^{(m+1)}=\sum_{n{\rm{ = }}1}^{N} \left(R{_n^{\mathrm{off}}}(\mathbf{a}^{(m+1)})+R{_n^{\mathrm{loc}}}(a_n^{(m+1)})\right)$; \\
      ~\quad c) $m=m+1$; \\
\STATE~~\textbf{End Repeat 3} until convergence, i.e., $|R_3^{(m)}-R_3^{(m-1)}|<\varepsilon_3$ \\
      ~~($m>1$), and obtain $\mathbf{a}_{\chi}=\mathbf{a}^{(m)}$;\\
\STATE Calculate the TCTB at the $\chi$-th iteration, which is denoted as $R_{\chi}=\sum_{n{\rm{ = }}1}^{N} \left(R{_n^{\mathrm{off}}}(\mathbf{a}_{\chi},\mathbf{w}_{n,\chi},\boldsymbol{\phi}_{\chi})+
R{_n^{\mathrm{loc}}}(a_{n,\chi})\right)$ by substituting $\mathbf{W}_\chi$, $\mathbf{a}_{\chi}$ and $\boldsymbol{\phi}_{\chi}$ into the objective function of problem (P0).\\
\STATE \textbf{End Repeat} until convergence, i.e., $|R{_{\chi}}-R{_{\chi-1}}|<\varepsilon$ ($\chi>1$), and obtain the maximum TCTB $R{_\chi}$ with the solution $\mathbf{W}^*=\mathbf{W}_{\chi}$, $\mathbf{a}^*=\mathbf{a}_{\chi}$, $\boldsymbol{\phi}^*=\boldsymbol{\phi}_{\chi}$. \\
}}
\end{algorithmic}
\end{algorithm}

The proposed three-step BCD optimization algorithm for solving the original TCTB maximization problem (P0) in \eqref{eq:MSCT} is summarized in \textbf{Algorithm} \ref{algorithmic1}, through which problem (P0) can be  effectively solved with guaranteed convergence \cite{B_Boyd04Convex}. In fact, the convergence can be easily proved based on Lemma \ref{lemma3}, \ref{lemma1}, and \ref{lemma2}, and we can show that the objective function of problem (P0) in \eqref{eq:MSCT_0} gradually increases with the iteration index $\chi$ through updating $\boldsymbol{\phi}$, $\mathbf{W}$, and $\mathbf{a}$ iteratively.

The computational complexity of the proposed three-step BCD optimization Algorithm \ref{algorithmic1} in each iteration mainly lies in the DC programming for solving problem (P1) to design the RIS reflecting coefficients and problem (P3) to optimize the energy partition parameters. Note that the complexity of solving problem (P1.1) in \eqref{eq:MSCTPhi_DCapp} and problem (P3.1) in \eqref{eq:MSCTaDC} can be estimated as with the order of $\mathcal{O}(K^6)$ and $\mathcal{O}(N^{3.5})$ according to the complexity of solving convex problems with interior point method\cite{TSP_K.Wang2014Outage}. Denote the number of iterations for solving problem (P1) and problem (P3) as $L_1$ and $L_3$, respectively, thus the total computational complexity of  Algorithm \ref{algorithmic1} can be further given as $O(L(L_1K^6+L_3N^{3.5}))$, where $L$ represents the total iteration number of the the BCD algorithm.
It is easy to observe that the complexity of the proposed algorithm increases dramatically  with the number of RIS reflecting elements and the number of UEs.

\section{Deep Learning Architectures}\label{sec:SL}
The proposed BCD Algorithm \ref{algorithmic1}  provides an effective optimization method for solving the TCTB maximization roblem (P0) in an iterative way. Although the BCD optimization algorithm can achieve effective solutions with guaranteed convergence, its high computational complexity may hinder it from being applied in real-time applications, which is a major bottleneck for most of the iterative optimization algorithms proposed in existing works \cite{JSAC_T.Bai2020Latency,ArXiv_S.Hua2019Reconfigurable,ArXiv_Y.Liu2020Intelligent,ArXiv_T.Bai2020Resource}. However, the effectiveness, robustness, and computational overhead of online implementations are known as crucial indicators for  practical networks.
One way to overcome this drawback is leveraging the deep learning methods, not only due to the fact that DNNs are regarded as universal function approximators but also because deep learning is well known as a promising way to achieve effective online implementations \cite{J_K.Hornik1989Multilayer,B_I.Goodfellow2016Deep,TCCN_T.OShea2017AnIntroduction}. Hence, in this section, we explore the potentials of deep learning approaches in obtaining effective solutions of the original  problem (P0).

The proposed deep learning methods in this section  aim at reducing the computational complexity of the proposed BCD optimization algorithm by effectively emulating this algorithm via supervised learning, so as to  facilitate lightweight online implementation of the BCD algorithm with high accuracy and stability.
Moreover, the use of DNNs enabled the design of a location-only deep learning approach that reduces overheads for CSI estimation and feedback compared to the BCD algorithm.
Through training the constructed DNNs offline with data samples generated from the BCD algorithm, the DNNs are capable of learning the inherent mappings of the algorithm and output effective solutions  mimicking this algorithm. Hence, we can use the trained DNNs to predict the required solutions online with significantly reduced computational complexity/running time.
Specifically, we resort to the deep learning approaches to obtain partial solution of problem (P0), including the RIS reflecting coefficients in $\boldsymbol{\phi}$ and the energy partition parameters in $\mathbf{a}$. Then,  we can directly obtain the receive beamforming vectors in $\mathbf{W}$ via Lemma \ref{lemma1} without learning.
In this way, we can effectively combine the prior knowledge (Lemma \ref{lemma1}) with deep learning to achieve the required solution  of the proposed BCD algorithm, with reduced cost for constructing, training, and testing the DNNs.

\begin{figure}[htbp] 
\centering
\includegraphics[scale=0.42]{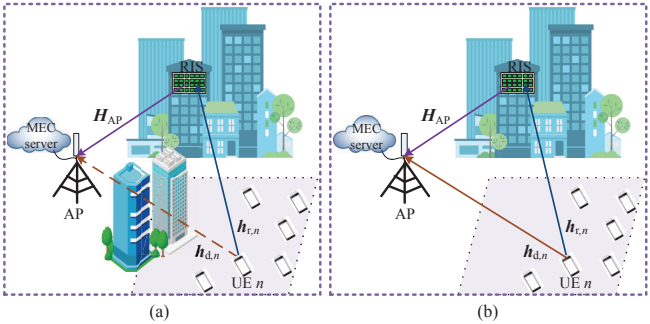}
\vspace{-2mm}
\caption{Two typical RIS-aided edge computation offloading scenarios in terms of whether LoS direct links exist between UEs and the AP: (a) Scenario without LoS direct links between UEs and AP where the LoS direct paths are blocked by objects such as buildings (urban area); (b) Scenario with strong LoS direct links between UEs and AP where no obstruction exists along the direct signal paths (suburb area).}
\label{SystemModel_NLOS_LOS}
\end{figure}

As mentioned before, the channel links between UEs and the RIS as well as that between the RIS and AP are very likely to be LoS channels, when the location of the RIS is carefully planned. For the multiplexing computation offloading mechanism considered in this paper, the direct links between UEs and the AP play a significant role in providing multi-path diversity gain for wireless communications. Two typical  RIS-aided edge computation offloading scenarios in terms of  whether LoS direct links exist between UEs and AP are considered here as shown in Fig.~\ref{SystemModel_NLOS_LOS}, where one is without LoS direct links denoted as scenario (a) and the other is with strong LoS direct links denoted as scenario (b).
In order to effectively deal with these two scenarios, we construct two deep learning architectures  in the following two subsections.
\vspace{-2mm}

\vspace{-4mm}
\subsection{CSI-Based Deep Learning Architecture}\label{sec:SL-Channel}
\begin{figure}[htbp]
\centering
\includegraphics[scale=0.6]{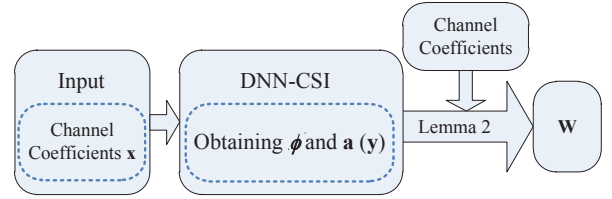} 
\vspace{-2mm}
\caption{The architecture for obtaining the  solutions of $\{\boldsymbol{\phi},\mathbf{a},\mathbf{W}\}$  with the CSI-based DNN-CSI.}
\label{DNN_Channel}
\end{figure}

For scenario (a) without LoS direct links between UEs and AP, a CSI-based deep learning architecture is given, as shown in Fig.~\ref{DNN_Channel}, to obtain the solutions of $\{\boldsymbol{\phi},\mathbf{a},\mathbf{W}\}$.
The real and imaginary parts of the channel coefficients $\{\mathbf{h}_{\mathrm{d},n}\}$,  $\{\mathbf{h}_{\mathrm{r},n}\}$ and $\mathbf{H}_{\mathrm{AP}}$ constitute the  input feature of the constructed DNN-CSI, represented by the input vector $\mathbf{x}$ with a dimension of $I=2(MN+KN+MK)$.
In contrast, the normalized angles of the RIS reflecting coefficients $\boldsymbol{\phi}$, denoted as $\boldsymbol{\widetilde{\theta}}=\boldsymbol{\theta}/2\pi$, and the energy partition parameters in $\mathbf{a}$ constitute the corresponding output vector $\mathbf{y}=[\widetilde{\theta}_1,\dots,\widetilde{\theta}_K,a_1,\dots,a_N]$ with the dimension of $(K+N)$.  It is easy to note that all the elements of the output vector are within the range of $[0,1]$,
and thus we can use the sigmoid function, i.e., $\mathrm{Sigmoid}(z)=\frac{1}{1+e^{-z}}$ as the output activation function.
With the final output $\boldsymbol{\phi}$ and $\mathbf{a}$ of the DNN-CSI, the optimal receive beamforming vector for each UE $n\in\mathcal{N}$, i.e., $\mathbf{w}_n$, can be readily obtained according to Lemma~\ref{lemma1} in Section \ref{ReceiveBeamformingDesign}.
\vspace{-4mm}

\vspace{-4mm}
\setlength{\tabcolsep}{0.3 pt}\begin{table}[thb]
\centering
\caption{Layout of  DNN-CSI for Obtaining $\boldsymbol{\phi}$ and $\mathbf{a}$}\label{table_DL_channel}
\vspace{-2mm}
\begin{tabular}{l|c|c|c}
\hline
~~\textbf{Layer}&~{\textbf{Size}} &~{\textbf{Parameter}} &~{\textbf{Activation}} \\
\hline
~~Input Layer~          &~$I$~ &~-~         &~-~\\
~~Layer1-1 (Dense)~       &~1024~        &~1024($I$+1)~ &~ELU~\\ 
~~Layer1-2 (BN)~          &~1024~        &~4096~      &~-~\\
~~Layer1-3 (Dropout 0.1)~ &~1024~        &~0~         &~-~\\
~~Layer2-1 (Dense)~       &~512~         &~524800~    &~ELU~\\
~~Layer2-2 (BN)~          &~512~         &~2048~      &~-~\\
~~Layer2-3 (Dropout 0.1)~ &~512~         &~0~         &~-~\\
~~Layer3-1 (Dense)~       &~256~         &~131328~    &~ELU~\\
~~Layer3-2 (BN)~          &~256~         &~1024~      &~-~\\
~~Layer3-3 (Dropout 0.1)~ &~256~         &~0~         &~-~\\
~~Layer4-1 (Dense)~      &~128~         &~32869~     &~ELU~\\
~~Layer4-2 (BN)~         &~128~         &~512~       &~-~\\
~~Layer4-3 (Dropout 0.1)~&~128~         &~0~         &~-~\\
~~Layer5-1 (Dense)~      &~128~         &~16512~     &~ELU~\\
~~Layer5-2 (BN)~         &~128~         &~512~       &~-~\\
~~Layer5-3 (Dropout 0.05)~&~128~         &~0~         &~-~\\
~~Output Layer (Dense)~  &~$K$+$N$~         &~($K$+$N$)(128+1)~     &~sigmoid~\\
\hline
~~\textbf{Total Trainable Parameters}~  &\multicolumn{3}{c}{1,632,288 ($M$=8,$N$=8,$K$=24)} \\
\hline
\end{tabular}
\end{table}

Here, we adopt a feedforward DNN with the layout in Table \ref{table_DL_channel}, which consists of an $I$-dimensional input layer, 5 normal hidden dense layers (layers 1-1, 2-1, 3-1, 4-1, 5-1), and a $(K+N)$-dimensional output layer, which are the key functional layers of DNN-CSI.
There are respectively 1024, 512, 256, 128, 128 neurons for the five hidden layers of the DNN-CSI.
The function of exponential linear units (ELU) is leveraged as the activation functions of the hidden layers with
\begin{align}
\mathrm{ELU}(z)\hspace{-4mm}&&&=\left\{
\begin{aligned}
& z, &&\mathrm{if} \ \ z>0, \\
& \alpha(\mathrm{exp}(z)-1), &&\mathrm{otherwie}, \ z\leq0,  \label{eq:ELU_fun}
\end{aligned}\right.
\end{align}
which has many attractive advantages such as high learning speed, high robustness with zero-centered outputs, etc.\footnote{We use the default form of ELU function in the Keras platform with $\alpha$=1.}
It should be noted that we add the Batch-Normalization layers (layers 1-2, 2-2, 3-2, 4-2, 5-2) and Dropout layers (layers 1-3, 2-3, 3-3, 4-3, 5-3) between two normal dense layers to accelerate the training speed, avoid gradients vanishing, as well as prevent overfitting of the DNN. To be specific, this fully connected feedforward DNN-CSI is with 10$\%$ of random dropout of neurons for the hidden layer 1 to hidden layer 4 and 5$\%$ of random dropout for the hidden layer 5 during each training epoch, so as to avoid overfitting.

\vspace{-2mm}
\subsection{Location-Only Deep Learning Architecture}\label{sec:SL-Location}
For scenario (b) with strong LoS direct links between UEs and AP, the CSI-based  deep learning architecture given in the previous subsection is still applicable. Nevertheless, note that the channel coefficients as well as the solutions of $\{\boldsymbol{\phi},\mathbf{a},\mathbf{W}\}$ are highly related to the locations of UEs, i.e., $\{(x_n,y_n)\}_{n\in\mathcal{N}}$, in this scenario, and thus we may use the UEs' locations as the only input feature of DNNs to obtain  $\{\boldsymbol{\phi},\mathbf{a},\mathbf{W}\}$.
In our considered scenarios, we assume that each UE is installed with an advanced global positioning system (GPS) module for outdoor  localization \cite{GPS_GOV2020} and is capable to apply the Wi-Fi round-trip time technology and standards for indoor localization \cite{Google2018_F.Diggelen_How}, through which UEs' location information can be obtained with  high accuracy.\footnote{According to the data shown in the website of GPS.gov, for high-end users with dual-frequency receivers and/or augmentation systems, the GPS accuracy can be dramatically boosted, which can enable real-time positioning within a few centimeters and long-term measurements at the millimeter level \cite{GPS_GOV2020}. It is shown in  \cite{Google2018_F.Diggelen_How} that an one-meter accuracy indoor localization is available  for  smart devices through the Wi-Fi round-trip time technology by 2018. } 

\begin{figure}[htbp]
\centering
\includegraphics[scale=0.6]{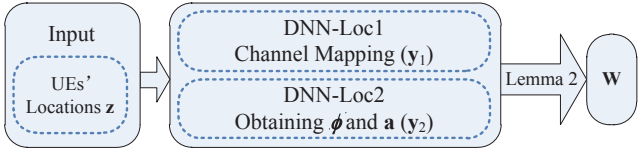}
\vspace{-2mm}
\caption{The architecture for obtaining the  solutions of $\{\boldsymbol{\phi},\mathbf{a},\mathbf{W}\}$  with the location-only DNN-Loc1 and DNN-Loc2.}
\label{DNN_Channel_Location}
\end{figure}

In order to further use the known relations between the solutions, e.g., Lemma~\ref{lemma1} presenting the relationship between $\{\mathbf{W}\}$ with $\{\boldsymbol{\phi},\mathbf{a}\}$  and the channel coefficients, a location-only deep learning architecture is proposed as shown in Fig.~\ref{DNN_Channel_Location}.
Here, two DNNs are constructed, where DNN-Loc1 aims at calculating the channel mapping between the UEs' locations and the channel coefficients with 2$N$-dimensional input feature $\mathbf{z}$ and $I$-dimensional output feature denoted as $\mathbf{y}_1$ while DNN-Loc2 focuses on obtaining $\{\boldsymbol{\phi},\mathbf{a}\}$ with input feature $\mathbf{z}$  and $(K+N)$-dimensional output feature denoted as $\mathbf{y}_2$. Then the optimal receive beamforming matrix, i.e., $\mathbf{W}$, can be easily calculated based on Lemma~\ref{lemma1} in Section \ref{ReceiveBeamformingDesign}. Note that the complicated pilot channel estimation and feedback can be removed when utilizing the location-only deep learning architecture for online implementation.
\vspace{-2mm}

\vspace{-4mm}
\setlength{\tabcolsep}{0.3 pt}\begin{table}[thb]
\centering
\caption{Layout of DNN-Loc1 for Channel Mapping}\label{table_DL_location_Chan}
\vspace{-2mm}
\begin{tabular}{l|c|c|c}
\hline
~~\textbf{Layer}&~{\textbf{Size}} &~{\textbf{Parameter}} &~{\textbf{Activation}} \\
\hline
~~Input Layer~          &~2$N$~        &~-~         &~-~\\
~~Layer1-1 (Dense)~       &~512~         &~512(2$N$+1)~ &~ELU~\\ 
~~Layer1-2 (BN)~          &~512~         &~2048~      &~-~\\
~~Layer1-3 (Dropout 0.1)~ &~512~         &~0~         &~-~\\
~~Layer2-1 (Dense)~       &~512~         &~262656~    &~ELU~\\
~~Layer2-2 (BN)~          &~512~         &~2048~      &~-~\\
~~Layer2-3 (Dropout 0.1)~ &~512~         &~0~         &~-~\\
~~Layer3-1 (Dense)~       &~256~         &~131328~    &~ELU~\\
~~Layer3-2 (BN)~          &~256~         &~1024~      &~-~\\
~~Layer3-3 (Dropout 0.1)~ &~256~         &~0~         &~-~\\
~~Layer4-1 (Dense)~      &~128~         &~32869~     &~ELU~\\
~~Layer4-2 (BN)~         &~128~         &~512~       &~-~\\
~~Layer4-3 (Dropout 0.1)~&~128~         &~0~         &~-~\\
~~Layer5-1 (Dense)~      &~256~         &~33024~     &~ELU~\\
~~Layer5-2 (BN)~         &~256~         &~1024~       &~-~\\
~~Layer5-3 (Dropout 0.05)~&~256~        &~0~         &~-~\\
~~Output Layer (Dense)~  &~$I$~        &~$I$(256+1)~   &~sigmoid~\\
\hline
~~\textbf{Total Trainable Parameters}~  &\multicolumn{3}{c}{702,208 ($M$=8,$N$=8,$K$=24)}\\
\hline
\end{tabular}
\end{table}

The layout of the feedforward DNN-Loc1 and DNN-Loc2 are given  in Table \ref{table_DL_location_Chan} and \ref{table_DL_location_aRIS}, respectively, where both have 5 normal hidden dense layers.
There are respectively 512, 512, 256, 128, 256 neurons for the five hidden layers of the DNN-Loc1, and respectively 512, 256, 128, 64, 32 neurons for the five hidden layers of the DNN-Loc2.
Similarly, the layers of Batch-Normalization and Dropout are also utilized for the DNN-Loc1 and DNN-Loc 2 with same dropout police of the DNN-CSI. Here, the sigmoid activation function is not only leveraged at the output layer of the DNN-Loc2 for obtaining $\{\boldsymbol{\phi},\mathbf{a}\}$ but also at that of the DNN-Loc1 for channel mapping where the output data samples 
are scaled into the range of $[0,1]$ with the MinMaxScaler in Tensorflow. In the testing stage, an inverse transformation of MinMaxScaler is used to achieve the required form of the output feature.
\vspace{-2mm}
\setlength{\tabcolsep}{0.3 pt}\begin{table}[thb]
\centering
\caption{Layout of DNN-Loc2 for Obtaining $\boldsymbol{\phi}$ and $\mathbf{a}$}\label{table_DL_location_aRIS}
\vspace{-2mm}
\begin{tabular}{l|c|c|c}
\hline
~~\textbf{Layer}&~{\textbf{Size}} &~{\textbf{Parameter}} &~{\textbf{Activation}} \\
\hline
~~Input Layer~          &~2$N$~       &~-~         &~-~\\
~~Layer1-1 (Dense)~       &~512~        &~512(2$N$+1)~ &~ELU~\\ 
~~Layer1-2 (BN)~          &~512~        &~2048~      &~-~\\
~~Layer1-3 (Dropout 0.1)~ &~512~        &~0~         &~-~\\
~~Layer2-1 (Dense)~       &~256~        &~131328~    &~ELU~\\
~~Layer2-2 (BN)~          &~256~        &~1024~      &~-~\\
~~Layer2-3 (Dropout 0.1)~ &~256~        &~0~         &~-~\\
~~Layer3-1 (Dense)~       &~128~        &~32896~     &~ELU~\\
~~Layer3-2 (BN)~          &~128~        &~512~       &~-~\\
~~Layer3-3 (Dropout 0.1)~ &~128~        &~0~         &~-~\\
~~Layer4-1 (Dense)~      &~64~         &~8256~      &~ELU~\\
~~Layer4-2 (BN)~         &~64~         &~256~       &~-~\\
~~Layer4-3 (Dropout 0.1)~&~64~         &~0~         &~-~\\
~~Layer5-1 (Dense)~      &~32~         &~2080~      &~ELU~\\
~~Layer5-2 (BN)~         &~32~         &~128~       &~-~\\
~~Layer5-3 (Dropout 0.05)~&~32~        &~0~         &~-~\\
~~Output Layer (Dense)~  &~$K$+$N$~     &~($K$+$N$)(32+1)~   &~sigmoid~\\
\hline
~~\textbf{Total Trainable Parameters}~  &\multicolumn{3}{c}{186,304 ($M$=8,$N$=8,$K$=24)}\\
\hline
\end{tabular}
\end{table}

\vspace{-4mm}
\subsection{Input Feature Uncertainty}\label{sec:DL_IFU}
In the previous subsection, an ideal scenario is considered where we assume that the input features to the CSI-based and the location-only DNNs, i.e.,  the input vector $\mathbf{x}$ of CSI in Fig. \ref{DNN_Channel} and the input vector $\mathbf{z}$ of UEs' locations in Fig. \ref{DNN_Channel_Location}, are perfectly known. In this case, the constructed DNNs can be  trained and tested based on the perfect input information of CSI and UEs' locations. However, the obtained CSI and UEs' locations are usually imperfect in practice due to the deviation of channel estimation and GPS/Wi-Fi localization.

In this section, we focus on a more practical scenario where the input features of CSI and UEs' locations for the CSI-based DNN and the location-only DNNs are corrupted with  uncertainty.
For the input vector $\mathbf{x}$ of CSI, the corresponding corrupted  counterpart is $\mathbf{\hat{x}}=\mathbf{x}+\triangle\mathbf{x}$, where $\triangle\mathbf{x}\sim\mathcal{N}(0,\mathbf{\sigma}{_{\triangle\mathbf{x}}^2})$ following the normal distribution is the random offset of the achieved CSI to the perfect CSI.
For the input vector $\mathbf{z}$ of UEs' locations, the corresponding corrupted counterpart is $\mathbf{\hat{z}}=\mathbf{z}+\triangle\mathbf{z}$,
where $\triangle\mathbf{z}\sim\mathcal{N}(0,\mathbf{\sigma}{_{\triangle\mathbf{z}}^2})$  is the random  offset (in meter) of the achieved UEs' locations to the perfect ones.\footnote{In the simulation results, the default standard deviation of  $\mathbf{\sigma}_{\triangle\mathbf{x}}$ is set as 0.001 according to the setting in \cite{ICASSP2020_F.Sohrabi_Robust} and the default standard deviation of $\mathbf{\sigma}_{\triangle\mathbf{z}}$ is set as 1  based on the localization accuracy of GPS given in GPS.gov \cite{GPS_GOV2020} and the Wi-Fi round-trip time technology shown in \cite{Google2018_F.Diggelen_How}.}  

In this practical case with uncertain input features, the CSI-based DNN and the location-only DNNs are trained and tested based on the corrupted input information of CSI and UEs' locations, respectively. The effectiveness and robustness of the two proposed deep learning architectures are validated by comparing their performance in the cases with perfect and imperfect input features, also comparing with the BCD optimization algorithm, which will be shown in Section \ref{sec:simulation}.

\vspace{-2mm}
\section{Implementation and Complexity Reduction of the Deep Learning Approaches}\label{sec:DL_Comparison}
In this section, we indicate the implementation setting of the proposed CSI-based and location-only deep-learning architectures.  In addition, the comparison results of the two deep learning methods  as well as the proposed BCD optimization algorithm in terms of the average running time are given, which further validates the potentials of the two proposed data-driven  approaches in reducing the computational complexity for achieving lightweight online implementations.

The training (including validation) and testing of the constructed DNNs 
are implemented based on the platforms of Tensorflow and Keras via supervised learning.
Also, the adaptive moment estimation (Adam) optimizer is utilized to train the DNNs with adaptive learning rates.
We adopt the mean absolute error (MAE) as the loss function for the CSI-based DNN-CSI given in Table \ref{table_DL_channel} and the location-only DNN-Loc2 given in Table \ref{table_DL_location_aRIS} for obtaining $\boldsymbol{\phi}$ and $\mathbf{a}$.
Through training the weights and bias terms between layers, the input-output mappings of these two DNNs are  driven to emulate the inherent mapping of the proposed BCD optimization algorithm.
In contrast, the mean square error (MSE) is leveraged as the loss function for the location-only DNN-Loc1 given in Table \ref{table_DL_location_Chan} for achieving the location-channel mapping.
The other parameters relating to training and testing the constructed DNNs are given in the following Table \ref{table_DL_Comparision}.

\vspace{-2mm}
\setlength{\tabcolsep}{0.3 pt}\begin{table}[thb]
\centering
\caption{Parameters related to Training and Testing}\label{table_DL_Comparision}
\vspace{-2mm}
\begin{tabular}{l|l}
\hline
~~~~{\textbf{Parameter}}~         &~~~{\textbf{Values}}~~~~ \\
\hline
~~~~Number of training samples~~~~ &~~~200000~  \\
~~~~Number of testing samples~  &~~~10000~  \\
~~~~Batch size                  &~~~128~   \\
~~~~Number of epoches~          &~~~1000~   \\
~~~~Initial learning rate       &~~~0.001~  \\
~~~~Validation split            &~~~0.2~    \\
\hline
\end{tabular}
\end{table}

\vspace{-6mm}
\subsection{Comparison between Two Deep Learning Approaches}\label{sec:DL_CSI_Loc_Comparison}
For the CSI-based deep learning architecture, it is required to obtain CSI in advance via pilot channel estimation for time division duplex (TDD) systems.  Due to the random characteristics of wireless channels, it is quite difficult to estimate CSI accurately  especially considering the pilot contamination. Moreover, the difficulty for wireless channel estimation may dramatically increase and the accuracy may degrade when the adopted subcarriers are with higher frequencies or the APs are with larger set of antennas which lead to more random multi-path fading or larger dimension size of CSI. 

In comparison, UEs' location information is less random and its dimension size being independent of the number of APs' antennas is much smaller  than that of the corresponding CSI. Hence, it is much easier and more convenient to obtain UEs' location information in practice. In addition, highly accurate location information for UEs in both outdoor and indoor scenarios can be guaranteed thanks to the advanced GPS modules \cite{GPS_GOV2020} and the Wi-Fi round-trip time technology \cite{Google2018_F.Diggelen_How}.
In fact, the location-only deep learning architecture provides a promising way to emulate the proposed BCD algorithm for lightweight online implementation, which can remove the complicated pilot channel estimation and feedback prior to wireless communications for task offloading.

Furthermore, in Table \ref{table_DL_Comparision1}, we present the values of trainable parameters, training time, testing time, and average inference time of the constructed DNN-CSI, DNN-Loc1, DNN-Loc2 for the case with $M=8$, $N=8$, $K=24$ in both cases with  perfect and imperfect input features.\footnote{The training, testing, and inference time in Table  \ref{table_DL_Comparision1} correspond to the processing time by a computer with 64-bit Intel(R) Core(TM) i5-9600KF CPU @3.7GHz and 32 GB RAM, running Python 3.7.7 and Tensorflow 2.1.0. Note that the training and testing time can be further reduced when implemented through more powerful computing servers.}
It is shown that the time overhead for training and testing the constructed DNNs with imperfect input features is slightly larger than that with perfect input features due to the fact that more complicated input-output mappings need to be figured out.
Note that the total  required training parameters of the two location-only DNNs are considerably less (nearly half) than that of the DNN-CSI as shown in Table \ref{table_DL_channel}-\ref{table_DL_location_aRIS}.
Also, the training, testing, and inference overhead can be significantly reduced by leveraging the location-only data-driven method, which is verified by the much less required  training, testing, and average inference time of DNN-Loc1, DNN-Loc2 shown in Table \ref{table_DL_Comparision1} that are only around 60$\%$ of those for DNN-CSI in both cases.\footnote{The training, testing, and inference time of the location-only architecture are the maximum of those for DNN-Loc1 and DNN-Loc2 (such as \mbox{3.3504 h}, 0.2418 s, 24.18 $\mu$s for the case with perfect input feature) since these two DNNs can be trained and tested in a parallel way.}
Hence, it is of great benefits to leverage the location-only deep learning architecture in situations that it can provide satisfactory inference solutions, such as in the scenario with strong LoS direct links between UEs and AP.

\vspace{-2mm}
\setlength{\tabcolsep}{0.3 pt}\begin{table}[thb]
\centering
\caption{Processing Time of the Proposed Algorithms}\label{table_DL_Comparision1}
\vspace{-2mm}
\begin{tabular}{l|l|l|l}
\hline
~~{\textbf{Parameter}}~         &~{\textbf{DNN-CSI}}~  &~{\textbf{DNN-Loc1}}~ &~{\textbf{DNN-Loc2}}~~\\
\hline
~~Trainable parameters~  &~1,632,288~  &~702,208~  &~186,304~ \\
\hline
~~Training samples~  &~$(\mathbf{x},\mathbf{y})$~ &~$(\mathbf{z},\mathbf{y}_1)$~ &~$(\mathbf{z},\mathbf{y}_2)$~ \\
~~Training time~  &~5.7426 h~  &~3.3504 h~  &~1.3979 h~ \\
~~Testing time~   &~0.3883 s~  &~0.2418 s~  &~0.1025 s~ \\
~~Average inference time~   &~38.83 $\mu$s~  &~24.18 $\mu$s~ &~10.25 $\mu$s~ \\
\hline
~~Training samples~  &~$(\hat{\mathbf{x}},\mathbf{y})$~ &~$(\hat{\mathbf{z}},\mathbf{y}_1)$~ &~$(\hat{\mathbf{z}},\mathbf{y}_2)$~ \\
~~Training time~  &~6.0345 h~  &~3.5123 h~  &~1.4862 h~ \\
~~Testing time~   &~0.4015 s~  &~0.2568 s~  &~0.1156s~ \\
~~Average inference time~   &~40.15 $\mu$s~  &~25.68 $\mu$s~ &~11.56 $\mu$s~ \\ %
\hline
~~\textbf{Average  BCD Running Time}~  &\multicolumn{3}{c}{~28.7 s}\\
\hline
\end{tabular}
\end{table}

\vspace{-4mm}
\subsection{Complexity Reduction Compared with the BCD Algorithm}\label{sec:DL_BCD_Comparison}
As we mentioned in Section \ref{sec:SL}, the proposed deep learning methods aim at emulating the proposed BCD algorithm with reduced computational complexity, so as to make it possible for lightweight online implementation.
In Table \ref{table_DL_Comparision1}, the average running time of the BCD algorithm for one realization is also given for comparison, i.e., 28.7 s, which is almost $10^6$ times to those of the two proposed deep learning methods that the CSI-base and the location-only DNNs only require  38.83 $\mu$s (40.15 $\mu$s) and 24.18 $\mu$s (25.68 $\mu$s) for the case with perfect (imperfect) input features. This result effectively  validates the ability of the proposed deep learning architectures in reducing the computational complexity/running time for providing lightweight online inference solutions.
\vspace{-4mm}
\begin{figure}[htbp]
\centering
\includegraphics[scale=0.48]{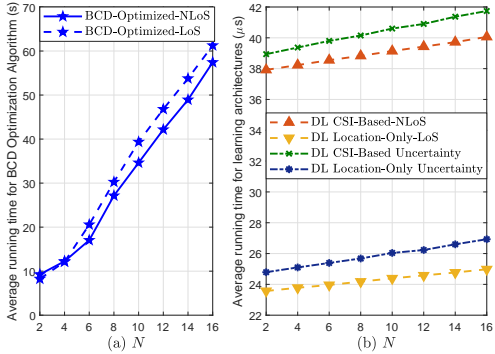} 
\vspace{-2mm}
\caption{The average running time for one realization of the proposed solutions versus the number of UEs ($N$) in the scenarios with NLoS and LoS direct links: (a) Optimization-based solutions, (b) Learning-based solutions without and with input uncertainty.}
\label{ImplementingTime_N}
\end{figure}

In Fig. \ref{ImplementingTime_N}, we further show more details of the average running time  for one realization of the proposed BCD optimization algorithm, the CSI-based and location-only deep learning architectures versus the number of UEs ($N$),  in the scenarios with NLoS and LoS direct links, respectively. In addition, the corresponding results of the two deep learning methods in the cases with input feature  uncertainty are also provided. It is noticeable that the average running time, i.e., the average inference time, of both two deep learning architectures are always quite small in all scenarios (measured in microsecond-$\mu$s) and  increase slightly with $N$, especially compared with those of the proposed BCD algorithm (measured in second-s) that increase quite considerably with $N$. Specifically, the average running time of the CSI-based leaning architecture is nearly millionth and the location-only learning architecture is less than millionth of that required by the corresponding BCD optimization solution, which indicates that lightweight online implementations are available by the proposed two data-driven architectures via periodically training.

\setlength{\tabcolsep}{0.3 pt}\begin{table*}[thb]
\centering
\caption{Simulation Parameters}\label{table1}
{\footnotesize{
\begin{tabular}{|l|l|l|}
\hline
~\textbf{Parameter }&~{\textbf{Symbol}} &~{\textbf{Value}} \\
\hline
~The parameters related to the square serving area\quad\quad\quad\quad\quad\quad\quad\quad~   &~$y_\mathrm{s}$, $D$ \quad\quad\quad\quad\quad\quad\quad\quad\quad &~20m, 40m \quad\quad\quad\quad\quad\quad \\
\hline
~The location of the AP~   &~$(0,y_{\mathrm{AP}},H_0)$  &~(0,20,5) m \\
\hline
~The location of the RIS~   &~$(x_\mathrm{R},0,H_\mathrm{R})$  &~(40,0,20) m \\
\hline
~The length of the time slot~   &~$T$  &~5 seconds \\
\hline
~Number of UEs &~$N$ &~8 \\
\hline
~Number of AP's antennas &~$M$ &~8  \\
\hline
~Number of RIS's reflecting elements &~$K=K_yK_z$ &~24=$8\times3$  \\
\hline
~Energy budgets of UEs &~$E_n~(n\in\mathcal{N})$ &~10 J\\
\hline
~Required CPU cycles per bit of UEs &~$C_n~(n\in\mathcal{N})$ &~200 cycles/bit \\
\hline
~The effective switched capacitance of UEs &~$\kappa_n (n\in\mathcal{N})$ &~$10^{-28}$ \\
\hline
~The total system bandwidth  &~$B$ &~40 MHz\\
\hline
~The noise power  &~$\sigma^2$ &~$-60$dBm \\
\hline
~The channel power gain at a reference distance of $d_0$=1 m &~$L_0$ &~$-10\mathrm{dB}$ \\
\hline
~The channel attenuation coefficients&~$\alpha_\mathrm{d}$, $\alpha_\mathrm{r}$, $\alpha_\mathrm{AP}$ &~3.5, 2.5, 2 \\
\hline
~The standard deviation of the offset to the perfect CSI &~$\mathbf{\sigma}_{\triangle\mathbf{x}}$ &~0.001 \\
\hline
~The standard deviation of the offset to the perfect UEs' locations &~$\mathbf{\sigma}_{\triangle\mathbf{z}}$ &~1 \\ 
\hline
\end{tabular}
}}
\end{table*}

\vspace{-2mm}
\section{Simulation Results}\label{sec:simulation}
In this section, simulation results are given to verify the effectiveness and performance improvement of the proposed BCD optimization algorithm as well as the CSI-based and location-only deep learning architectures.
In addition, the effectiveness and robustness of the  two proposed deep learning methods to the corrupted input features of CSI and UEs' locations with uncertainty is also validated by simulations.

A three-dimensional (3D) Euclidean coordinate system is adopted to describe the locations of the AP as $(0,y_\mathrm{AP},H)$, the RIS as $(x_\mathrm{R},0,H_\mathrm{R})$ and UE $n\in \mathcal{N}$ as $(x_n,y_n,0)$, all measured in meters (m) as shown in Fig. \ref{fig:system_model}.
The aided RIS is with a uniform rectangular array (URA) of $K=K_yK_z$ reflecting elements, while the $M$-antenna AP is equipped with a  uniform linear array (ULA).
We assume that the $N$ ground UEs are randomly distributed in a square serving area of $D\times D$ $\mathrm{m}^2$, with four vertices at horizontal locations of $(x_\mathrm{s},0)$, $(x_\mathrm{s}+D,0)$, $(x_\mathrm{s},D)$, and $(x_\mathrm{s}+D,D)$.
We consider the Rician fading channel model to account for both the LoS and non-LoS (NLoS) components of all the channels as  \cite{B_D.Tse2005Fundamentals}
\begin{align}
\mathbf{h}_{\mathrm{r},n}&=\sqrt{\frac{\varrho_\mathrm{r}}{1+\varrho_\mathrm{r}}}\mathbf{h}{_{\mathrm{r},n}^{\mathrm{LoS}}}+\sqrt{\frac{1}{1+\varrho_\mathrm{r}}}
\mathbf{h}{_{\mathrm{r},n}^{\mathrm{NLoS}}}, \ \forall n\in\mathcal{N}, \label{eq:h_rn}\\
\mathbf{H}_{\mathrm{AP}}&=\sqrt{\frac{\varrho_\mathrm{AP}}{1+\varrho_\mathrm{AP}}}\mathbf{H}{_{\mathrm{AP}}^{\mathrm{LoS}}}+\sqrt{\frac{1}{1+\varrho_\mathrm{AP}}}
\mathbf{H}{_{\mathrm{AP}}^{\mathrm{NLoS}}}, \label{eq:HAP}\\
\mathbf{h}_{\mathrm{d},n}&=\sqrt{\frac{\varrho_\mathrm{d}}{1+\varrho_\mathrm{d}}}\mathbf{h}{_{\mathrm{d},n}^{\mathrm{LoS}}}+\sqrt{\frac{1}{1+\varrho_\mathrm{d}}}
\mathbf{h}{_{\mathrm{d},n}^{\mathrm{NLoS}}}, \ \forall n\in\mathcal{N}, \label{eq:h_dn}
\end{align}
where $\varrho_\mathrm{r}$, $\varrho_\mathrm{AP}$, $\varrho_\mathrm{d}$ indicate the corresponding Rician factors. Without loss of generality, we denote  $\zeta_\mathrm{r}=\frac{\varrho_\mathrm{r}}{1+\varrho_\mathrm{r}}$, $\zeta_\mathrm{AP}=\frac{\varrho_\mathrm{AP}}{1+\varrho_\mathrm{AP}}$, $\zeta_\mathrm{d}=\frac{\varrho_\mathrm{d}}{1+\varrho_\mathrm{d}}$ as the Rician parameters related to the LoS components which are used to generate the channels in the simulations. Assuming that a half-wavelength spacing is assumed among adjacent elements/antennas at the RIS and AP, the LoS components modeled in the angular domain are then given as \cite{B_D.Tse2005Fundamentals,JOE_L.YuanGain2014Gain}
\begin{align}
\hspace{-4mm}\mathbf{h}{_{\mathrm{r},n}^{\mathrm{LoS}}}&=\sqrt{L_{\mathrm{r},n}}\mathbf{e}{_{\mathrm{r},n}^\mathrm{r}}
(\beta{_{\mathrm{r},n}^\mathrm{r}},\gamma{_{\mathrm{r},n}^\mathrm{r}}), \ \forall n, \label{eq:h_rn_LOS}\\
\hspace{-4mm}\mathbf{H}{_{\mathrm{AP}}^{\mathrm{LoS}}}&=\sqrt{L_{\mathrm{AP}}}\mathbf{e}{_{\mathrm{AP}}^\mathrm{r}}
(\beta{_{\mathrm{AP}}^\mathrm{r}})(\mathbf{e}{_{\mathrm{R}}^\mathrm{t}}(\beta{_{\mathrm{R}}^\mathrm{t}},\gamma{_{\mathrm{R}}^\mathrm{t}}))^\mathrm{H}, \label{eq:HAP_LOS}\\
\hspace{-4mm}\mathbf{h}{_{\mathrm{d},n}^{\mathrm{LoS}}}&=\sqrt{L_{\mathrm{d},n}}\mathbf{e}{_{\mathrm{d},n}^\mathrm{r}}
(\beta{_{\mathrm{d},n}^\mathrm{r}}), \ \forall n, \label{eq:h_dn_LOS}
\end{align}
where $\mathbf{e}{_{\mathrm{r},n}^\mathrm{r}}(\beta{_{\mathrm{r},n}^\mathrm{r}},\gamma{_{\mathrm{r},n}^\mathrm{r}})\in\mathbb{C}^{K\times1}=
\mathbf{e}{_{\mathrm{r},n,y}^\mathrm{r}}(\beta{_{\mathrm{r},n}^\mathrm{r}},\gamma{_{\mathrm{r},n}^\mathrm{r}})
\otimes\mathbf{e}{_{\mathrm{r},n,z}^\mathrm{r}}(\beta{_{\mathrm{r},n}^\mathrm{r}},\gamma{_{\mathrm{r},n}^\mathrm{r}})$ with
$\mathbf{e}{_{\mathrm{r},n,y}^\mathrm{r}}(\beta{_{\mathrm{r},n}^\mathrm{r}},\gamma{_{\mathrm{r},n}^\mathrm{r}})=
\{\mathrm{exp}(j\pi(k_y-1)$ $\mathrm{sin}\beta{_{\mathrm{r},n}^\mathrm{r}}\mathrm{sin}\gamma{_{\mathrm{r},n}^\mathrm{r}})\}
_{k_y=1}^{K_y}\in\mathbb{C}^{K_y\times1}$ and
$\mathbf{e}{_{\mathrm{r},n,z}^\mathrm{r}}(\beta{_{\mathrm{r},n}^\mathrm{r}},\gamma{_{\mathrm{r},n}^\mathrm{r}})=
\{\mathrm{exp}(j\pi(k_z-1)\mathrm{cos}\beta{_{\mathrm{r},n}^\mathrm{r}}\mathrm{sin}\gamma{_{\mathrm{r},n}^\mathrm{r}})\}
_{k_z=1}^{K_z}\in\mathbb{C}^{K_z\times1}$,
$\mathbf{e}{_{\mathrm{AP}}^\mathrm{r}}(\beta{_{\mathrm{AP}}^\mathrm{r}})=\{\mathrm{exp}(j\pi(m-1)\mathrm{sin}\beta{_{\mathrm{AP}}^\mathrm{r}})\}_{m=1}^M\in\mathbb{C}^{M\times1}$ and
$\mathbf{e}{_{\mathrm{d},n}^\mathrm{r}}(\beta{_{\mathrm{d},n}^\mathrm{r}})=\{\mathrm{exp}(j\pi(m-1)\mathrm{sin}\beta{_{\mathrm{d},n}^\mathrm{r}})\}_{m=1}^M\in\mathbb{C}^{M\times1}$ are the receive array steering vectors  with the effective angles of arrival (AOAs).
Also,
$\mathbf{e}{_{\mathrm{R}}^\mathrm{t}}(\beta{_{\mathrm{R}}^\mathrm{t}},\gamma{_{\mathrm{R}}^\mathrm{t}})\in\mathbb{C}^{K\times1}=
\mathbf{e}{_{\mathrm{R},y}^\mathrm{t}}(\beta{_{\mathrm{R}}^\mathrm{t}},\gamma{_{\mathrm{R}}^\mathrm{t}})
\otimes\mathbf{e}{_{\mathrm{R},z}^\mathrm{t}}(\beta{_{\mathrm{R}}^\mathrm{t}},\gamma{_{\mathrm{R}}^\mathrm{t}})$ is the transmit array steering vector with the effective angles of departure (AOD), where
$\mathbf{e}{_{\mathrm{R},y}^\mathrm{t}}(\beta{_{\mathrm{R}}^\mathrm{t}},\gamma{_{\mathrm{R}}^\mathrm{t}})=
\{\mathrm{exp}(j\pi(k_y-1)$ $\mathrm{sin}\beta{_{\mathrm{R}}^\mathrm{t}}\mathrm{sin}\gamma{_{\mathrm{R}}^\mathrm{t}})\}
_{k_y=1}^{K_y}\in\mathbb{C}^{K_y\times1}$ and
$\mathbf{e}{_{\mathrm{R},z}^\mathrm{t}}(\beta{_{\mathrm{R}}^\mathrm{t}},\gamma{_{\mathrm{R}}^\mathrm{t}})=
\{\mathrm{exp}(j\pi$ $(k_z-1)\mathrm{cos}\beta{_{\mathrm{R}}^\mathrm{t}}\mathrm{sin}\gamma{_{\mathrm{R}}^\mathrm{t}})\}
_{k_z=1}^{K_z}\in\mathbb{C}^{K_z\times1}$. Here, $\beta$ and $\gamma$ respectively represent the elevation and azimuth of AOA or AOD.

$L_{\mathrm{r},n}$,  $L_{\mathrm{AP}}$ and  $L_{\mathrm{d},n}$ in \eqref{eq:h_rn_LOS}-\eqref{eq:h_dn_LOS} model the distance-dependent path loss of the corresponding channels.
Suppose that each element of the RIS has a 3 dBi gain due to the fact that only the  front half-space reflects signals  \cite{Arxiv_Q.Wu2019Joint}, 
while each antenna of the AP has an isotropic radiation pattern with 0 dBi antenna gain. Then we have $L_{\mathrm{r},n}=10^{0.3}L_0(d_{\mathrm{r},n}/d_0)^{-\alpha_\mathrm{r}}$, $L_{\mathrm{AP}}=10^{0.3}L_0(d_{\mathrm{AP}}/d_0)^{-\alpha_\mathrm{AP}}$ and $L_{\mathrm{d},n}=L_0(d_{\mathrm{d},n}/d_0)^{-\alpha_\mathrm{d}}$,  where $d_{\mathrm{r},n}$, $d_{\mathrm{AP}}$, $d_{\mathrm{d},n}$ are the corresponding Euclidean distances between the transceivers, $L_0$ is the average constant  path loss for all the channels at the reference distance of $d_0$, and $\alpha_{\mathrm{r}}$, $\alpha_{\mathrm{AP}}$, $\alpha_{\mathrm{d}}$ are the channel attenuation coefficients.

In addition, the NLoS components of the channels in \eqref{eq:h_rn}-\eqref{eq:h_dn} are modeled as the Rayleigh fading combining with the distance-dependent path loss as follows
\vspace{-2mm}
\begin{align}
\hspace{-2mm}\mathbf{h}{_{\mathrm{r},n}^{\mathrm{NLoS}}}&=\sqrt{L_{\mathrm{r},n}}\boldsymbol{\eta}_{\mathrm{r},n}
=\sqrt{10^{0.3}L_0(d_{\mathrm{r},n}/d_0)^{-\alpha_\mathrm{r}}}\boldsymbol{\eta}_{\mathrm{r},n},\label{h_relay_NLOS}\\
\hspace{-2mm}\mathbf{H}{_{\mathrm{AP}}^{\mathrm{NLoS}}}&=\sqrt{L_{\mathrm{AP}}}\boldsymbol{\Gamma}_{\mathrm{AP}}
=\sqrt{10^{0.3}L_0(d_{\mathrm{AP}}/d_0)^{-\alpha_\mathrm{AP}}}\boldsymbol{\Gamma}_{\mathrm{AP}},\label{h_AP_NLOS}\\
\hspace{-2mm}\mathbf{h}{_{\mathrm{d},n}^{\mathrm{NLoS}}}&=\sqrt{L_{\mathrm{d},n}}\boldsymbol{\eta}_{\mathrm{d},n}
=\sqrt{L_0(d_{\mathrm{d},n}/d_0)^{-\alpha_\mathrm{d}}}\boldsymbol{\eta}_{\mathrm{d},n}, \label{h_direct_NLOS}
\end{align}
where $\boldsymbol{\eta}_{\mathrm{r},n}\sim\mathcal{CN}(0,\mathbf{I}_K)$, $\boldsymbol{\Gamma}_{\mathrm{AP}}\sim\mathcal{CN}(0,\mathbf{I}_M)$, and $\boldsymbol{\eta}_{\mathrm{d},n}\sim\mathcal{CN}(0,\mathbf{I}_M)$ denote the corresponding Rayleigh fading coefficients.
In the following simulation results, we assume that the LoS channels between the RIS and the AP as well as UEs are achieved by deploying the RIS at a desirable location, and thus $\zeta_\mathrm{r}=\zeta_\mathrm{AP}=1$.
The other basic simulation parameters are listed in Table~\ref{table1} unless specified otherwise.

\vspace{-3mm}
\subsection{Results in the Scenario without LoS Direct Links}\label{sec:SR_SL_Channel}
\vspace{-1mm}
In this subsection, the LoS paths of UEs' direct links are blocked as shown in the scenario (a) of Fig. \ref{SystemModel_NLOS_LOS}, which is quite common in practical communications especially in central building districts of urban areas. Hence, we set $\zeta_\mathrm{d}=0$ in this subsection. Numerical results for the proposed optimization solution (`BCD-Optimized Solution'), the CSI-based deep learning solution (`DL CSI-Based') as well as its counterpart with input uncertainty (`DL CSI-Based Uncertainty') are presented in comparison with three traditional benchmarks, where the `Direct Offloading-No RIS' scheme corresponds to the case without deploying RIS, the `ZF Receive Beamforming' scheme considers the ZF beamforming for detecting UEs' signals as in Section \ref{ZF_Beamforming},  and the `Equal Energy Allocation' scheme is operated by equally allocating the UEs' energy budgets for local computing and computation offloading.\footnote{Note that the learning-based works mentioned in the Introduction section either focus on traditional MEC or RIS-aided downlink architectures, and none of them consider the RIS-aided MEC systems. Even though some optimization schemes of RIS-aided MEC systems were given in \cite{JSAC_T.Bai2020Latency,ArXiv_S.Hua2019Reconfigurable,ArXiv_Y.Liu2020Intelligent,ArXiv_T.Bai2020Resource}, different performance metrics or scenarios were considered as discussed in the Introduction. Hence, these works and our current work are not comparable.}

\begin{figure}[htbp]
\centering
\includegraphics[scale=0.48]{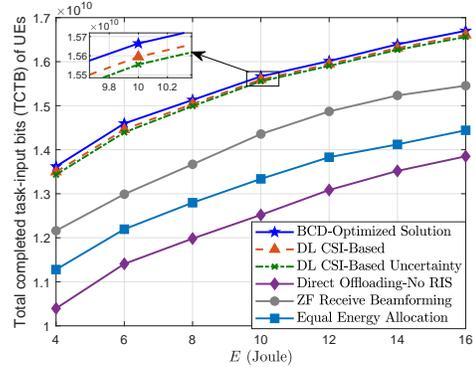} 
\vspace{-2mm}
\caption{The TCTB of UEs versus the UEs' uniform energy budget $E=E_n$ for $n\in\mathcal{N}$.}
\label{TCTB_RayRic0_E}
\end{figure}

In Fig. \ref{TCTB_RayRic0_E}, we first show the TCTB of all the considered schemes w.r.t. the UEs' uniform energy budget, i.e., $E=E_n$ for $n\in\mathcal{N}$. From this figure, we can observe that the TCTB curves of all the schemes increase with $E$, which coincides with the intuition that more computation task-input bits can be completed if the UEs are endowed with more energy. It is clear to see that significant performance improvement can be achieved by the proposed BCD-Optimized Solution, verifying the great benefits of deploying the aided RIS, also jointly optimizing the RIS coefficients, the receive beamforming and the UEs' energy allocation. It is confirmed that the proposed BCD
algorithm provides 26$\%$ improvement in TCTB over the benchmark of direct offloading without the assistance of RIS.
More importantly, the CSI-based deep learning method can achieve a performance very close to the proposed optimization solution no matter with perfect or imperfect input feature of CSI, which clearly demonstrates that the CSI-based deep learning architecture proposed in Section \ref{sec:SL-Channel} can effectively emulate the proposed BCD optimization algorithm, with a much reduced online complexity and high robustness.
\vspace{-2mm}
\vspace{-2mm}
\begin{figure}[htbp]
\centering
\includegraphics[scale=0.48]{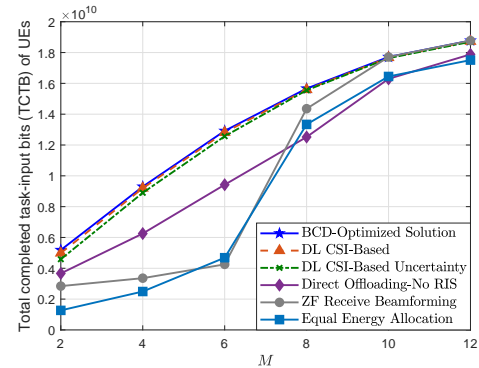}
\vspace{-2mm}
\caption{The TCTB of UEs versus the number of AP's antennas $M$ with $N=8$.}
\label{TCTB_RayRic0_M}
\end{figure}

The performance in terms of TCTB versus the number of the AP's antennas is presented in Fig. \ref{TCTB_RayRic0_M}. The effectiveness, robustness, and generalizability of the proposed CSI-Based deep learning architecture is further validated by the results that its performance can always approach that of the  BCD-Optimized Solution in both scenarios with and without input uncertainty no matter how many antennas are installed at the AP.  Although we can see that all the curves of TCTB increase as $M$ grows, it is obvious that the   performance of proposed optimization solution and the DL CSI-Based schemes are far more superior and stable than that of the other baseline solutions especially in the situations of $M< N$. When the AP has to serve more UEs than its installed antennas, the performance gap between the DL CSI-Based with and without input uncertainty is slightly larger, while the performances of the schemes with ZF Receive Beamforming and Equal Energy Allocation degrade dramatically. This is due to the fact that the ZF receive beamforming is incapable of separating out the signal streams when they are more than the number of receive antennas. Also, in these situations, the interference management through designing the UEs' energy allocation plays a significant role in guaranteeing the system performance.

\begin{figure}[htbp]
\centering
\includegraphics[scale=0.48]{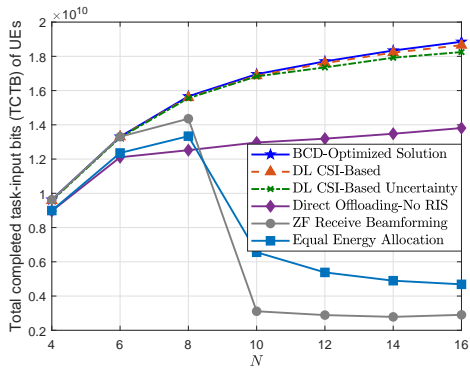}
\vspace{-2mm}
\caption{The TCTB of UEs versus the number of UEs $N$  with $M=8$.}
\label{TCTB_RayRic0_N}
\end{figure}

In Fig. \ref {TCTB_RayRic0_N}, we study the effects of the number of UEs, i.e., $N$, on the system performance of TCTB.  Here, the effectiveness  of the CSI-based deep learning architecture  is further verified in the scenarios  with different number of users considering both perfect and imperfect input CSI, which also demonstrates its robustness and the generalizability. Similar results can be observed as from Fig. \ref{TCTB_RayRic0_M} that our proposed BCD-Optimized Solution as well as the CSI-based deep learning schemes with and without input uncertainty have strong robustness in dealing with the cases when serving more UEs than the number of the AP's antennas. These cases are particularly relevant to massive connectivity scenarios. Instead of degrading the performance like the benchmarks, our proposed solutions are able to provide even better performance as $N$ becoming larger than $M$ through effectively designing the receive bearmforming vectors and the UE's energy allocation.
\vspace{-2mm}


\vspace{-6mm}
\subsection{Results in the Scenario with Strong LoS Direct Links}\label{sec:SR_SL_Location}
In this subsection, we focus on implementing the mentioned schemes in the scenario where strong LoS direct links exist for UEs in the considered serving area, which is exactly the scenario (b) of Fig. \ref{SystemModel_NLOS_LOS}. This scenario is practically relevant when considering the suburb districts,  and we set $\zeta_\mathrm{d}=1$ in the following simulation results. In this scenario, the location-only deep learning architecture (`DL Location-Only') can be leveraged to mimic the mapping of the proposed BCD algorithm. In addition, the performance of its counterpart solution with uncertain input UEs' locations denoted as `DL Location-Only Uncertainty' is also given in this subsection.
\vspace{-2mm}
\begin{figure}[htbp]
\centering
\includegraphics[scale=0.48]{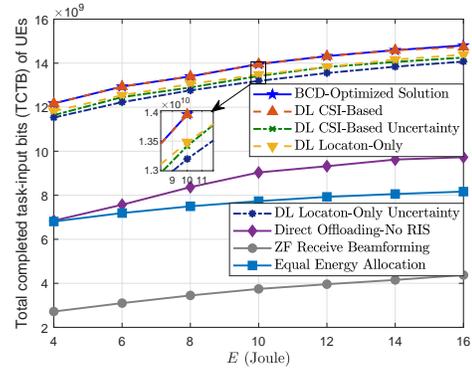}
\vspace{-2mm}
\caption{The TCTB of UEs versus the UEs' uniform energy budget $E=E_n$ for $n\in\mathcal{N}$.}
\label{TCTB_Ric10_E}
\end{figure}

We first show the TCTB performance of all the considered schemes versus the UEs' uniform energy budget $E$ in Fig. \ref{TCTB_Ric10_E}. Obviously, both the CSI-based and the location-only deep learning methods can achieve excellent system performance, in both scenarios with and without input uncertainty. Even though the DL Location-Only solution is slightly worse than the DL CSI-Based solution which is almost the same as the BCD-Optimized Solution, it is far more superior than the other benchmarks.
It is noticeable that the TCTB gap between the DL CSI-Based and DL Location-Only schemes becomes slightly smaller in the scenarios with uncertain input features. More importantly, the UEs' locations are quite easier to obtain compared with the related CSI, which makes it more flexible to achieve online implementation through the location-only deep learning architecture in both scenarios with perfect and imperfect input features. Also from this figure, we can clearly see that the ZF Receive Beamforming scheme is even worse than the scheme without RIS and the scheme of Equal Energy Allocation. The reason behind this is that in the considered  scenario with LoS direct links, the effective channels of different UEs may be highly correlated, and it is almost impossible to distinguish different UEs' data streams through ZF receive beamforming especially in the cases with $M\leq N$. This phenomenon further indicates the importance of effectively designing the AP's receive beamforming and managing the UEs' energy budgets.
\begin{figure}[htbp]
\centering
\includegraphics[scale=0.48]{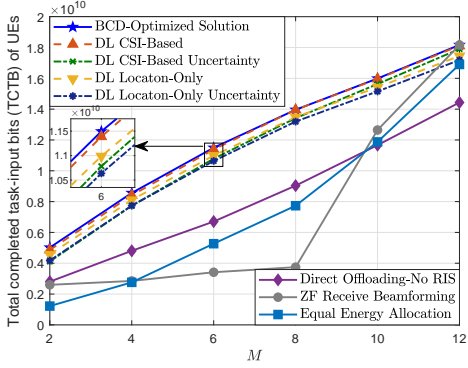}
\vspace{-2mm}
\caption{The TCTB of UEs versus the number of AP's antennas $M$ with $N=8$.}
\label{TCTB_Ric10_M}
\end{figure}

Fig. \ref{TCTB_Ric10_M} depicts the TCTB curves versus the number of the AP's antennas, i.e., $M$. Clearly, the results in this figure further demonstrate the effectiveness, the robustness and the generalizability of the CSI-based DNN architecture in different scenarios combined with the results in the previous subsection and the location-only DNN architecture in situations where AP is installed with different number of antennas.
If only uncertain input features are available, we can observe that the DL Location-only scheme can almost approach the DL CSI-Based when $M<N$.
Similar to the results in Fig. \ref{TCTB_Ric10_E}, the performance of the ZF Receive Beamforming scheme is unacceptable when $M\leq N$. When $M$ is greater than $N$, like $M=12$, the disadvantages of the scheme without RIS becomes more obvious since all the other schemes can achieve much better performance with the assistance of the RIS.
\vspace{-2mm}
\begin{figure}[htbp]
\centering
\includegraphics[scale=0.48]{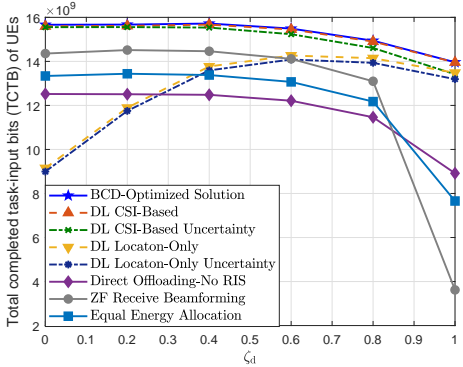} 
\vspace{-2mm}
\caption{The TCTB of UEs versus the Rician channel parameter $\zeta_\mathrm{d}$.}
\label{TCTB_Ric0246810_RicZeta}
\end{figure}

In Fig. \ref{TCTB_Ric0246810_RicZeta}, we  show the influence of the Rician channel parameter related to the LoS components of the direct links, i.e, $\zeta_\mathrm{d}$. It can be seen that the performance of the DL CSI-Based scheme can always achieve satisfactory performance very close to the proposed BCD-Optimized Solution if perfect CSI is available, no matter the  UEs' direct links  are highly faded without LoS components as in $\zeta_\mathrm{d}=0$, or with LoS components as $\zeta_\mathrm{d}>0$ and even $\zeta_\mathrm{d}=1$. This result indicates that the perfect input feature of CSI is capable of capturing sufficient information in emulating the proposed BCD optimized solution. In contrast, the DL Location-Only scheme can achieve good performance when $\zeta_\mathrm{d}$ is close to 1, but the performance degrades as $\zeta_\mathrm{d}$ decreases. This is reasonable since the optimization solution  becomes less relevant to the UE's locations as $\zeta_\mathrm{d}$ becomes smaller where the small-scale Rayleigh fading accounts more.
Compared with the DL CSI-Based  scheme, the performance of DL Location-Only solution is more stable and degrades slighter when $\zeta_\mathrm{d}=1$ with uncertain input feature.
Fig. \ref{TCTB_Ric0246810_RicZeta} verifies that the location-only DNN architecture is more suitable to the scenarios where strong LoS direct links for UEs are present, just coinciding with our original intention as in Section \ref{sec:SL-Location}.
Interestingly, we can see that the performance of the schemes except for the two DL Location-Only solutions degrades as $\zeta_\mathrm{d}$ increases, which highlights that the channel fading is beneficial to wireless communications when considering the multiplexing computation offloading since channel fading provides an additional degree of freedom for avoiding channel correlation \cite{B_D.Tse2005Fundamentals}. While for the case of $\zeta_\mathrm{d}=1$, the performance becomes worse since the UEs' channels are highly correlated and no  fading can be used to achieve the additional degree of freedom gain.
\vspace{-2mm}
\begin{figure}[htbp]
\centering
\includegraphics[scale=0.48]{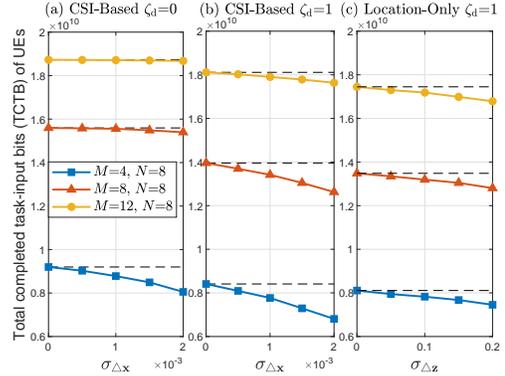}
\vspace{-2mm}
\caption{The TCTB of UEs versus the Input features' uncertainty level.}
\label{TCTB_sigma}
\end{figure}

Fig. \ref{TCTB_sigma} presents the TCTB performance of the two deep learning architectures versus the uncertainty levels of the CSI  and  UEs' locations, respectively represented by $\mathbf{\sigma}_{\triangle\mathbf{x}}$ and $\mathbf{\sigma}_{\triangle\mathbf{z}}$. Case (a) and (b) show the performance of the CSI-based learning solution in the scenarios with NLoS and LoS direct links, respectively, while case (c) demonstrates that of the location-only learning method in the scenario with LoS direct links. From these three cases, we can observe that the performance degradation is  much less when $M\geq N$, which is quite obvious for the CSI-based learning method in (a) and (b). In addition, the CSI-based learning solution degrades more serious in the scenario with LoS direct links in (b) compared with the NLoS case in (a), due to the fact that the DNN-CSI in LoS scenario (b) without channel fading is more sensitive to the input uncertainty in CSI. Based on (b) and (c), it is noticeable that the  location-only solution is less sensitive to the uncertain input feature in the scenario with LoS direct links, which further indicates its higher stability and robustness for fulfilling lightweight online implementation in this scenario.

\vspace{-2mm}
\section{Conclusion}\label{sec:conclusion}
In this paper, a RIS-aided MEC architecture with multiplexing computation offloading has been investigated, where the RIS constructively reflects the UEs' offloaded input-data-bearing signals to improve the UEs' computation efficiency.
During a given time slot, the TCTB of all the UEs with limited energy budgets is maximized by jointly optimizing the RIS reflecting coefficients, the receiving beamforming vectors and UEs' energy partition strategies for local computing and computation offloading. A three-step BCD optimization algorithm is proposed to solve the formulated non-convex TCTB maximization problem iteratively with guaranteed convergence. In addition, two deep learning architectures based on CSI and the UEs' locations are constructed to mimic the mapping of the BCD algorithm with a considerable complexity reduction.
The simulation results have confirmed that  significant performance improvement can be achieved by leveraging the proposed BCD algorithm comparing with some existing schemes. For both scenarios with perfect and imperfect input features, the CSI-based learning architecture can always approach the performance of the BCD algorithm, while the more practical location-only learning architecture can provide satisfactory and more robust  performance when strong LoS direct links exist between UEs and  AP.





\vspace{-1mm}
\bibliographystyle{IEEEtran}
\vspace{-2mm}
\bibliography{MEC_IRS_DL}

\end{document}